\documentclass[a4paper, fleqn]{cas-dc}

\usepackage[utf8]{inputenc}
\usepackage{amsmath}
\usepackage{amsthm}
\usepackage{amssymb}
\usepackage{bm}
\usepackage{graphicx}
\usepackage{longtable}
\usepackage{booktabs}
\usepackage{circuitikz}
\usepackage{svg}
\usepackage{pdfpages}
\usepackage{float}
\usepackage[inline,shortlabels]{enumitem}
\usepackage{stfloats}
\usepackage{supertabular}
\usepackage{booktabs}
\usepackage{caption}
\usepackage[numbers, sort&compress]{natbib}
\usepackage{flushend}
\usepackage{tikz}

\newcommand{\tikzcircle}[2][red,fill=red]{\tikz[baseline=-0.6ex]\draw[#1,radius=#2] (0,0) circle ;}%

\newlist{E}{enumerate}{1}
\setlist[E]{label=\textbf{E\arabic*:}}

\newlist{Q}{enumerate}{1}
\setlist[Q]{label=\textbf{Q\arabic*:}}

\newlist{Hlist}{enumerate}{1}
\setlist[Hlist]{label=\textbf{H\arabic*:}}

\usepackage[nameinlink]{cleveref}
\crefname{figure}{Fig.}{Figs.}
\Crefname{figure}{Figure}{Figures}
\crefname{equation}{Eq.}{Eqs.}
\Crefname{equation}{Equation}{Equations}
\crefname{section}{Section}{Sections}
\crefname{table}{Table}{Tables}
\crefname{appendix}{Appendix}{Appendices}

\newcommand{\eg}{e.g.,\ }

\newcommand{\cf}{cf.\ }
\newcommand{\ie}{i.e.,\ }

\newcommand{\centigrades}{{\textdegree{}C}}

\def\xbuild{x^\textrm{b}}
\def\xhatbuild{\hat{x}^\textrm{b}}
\def\xext{x^\textrm{e}}
\def\xhid{x^\textrm{h}}
\def\xobs{x^\textrm{obs}}
\def\xset{\textrm{X}}
\def\Xbuild{\xset^\textrm{b}}
\def\Xext{\xset^\textrm{e}}
\def\Xhid{\xset^\textrm{h}}
\def\Xobs{\xset^\textrm{obs}}
\def\X{\xset}
\def\U{\mathrm{U}}
\def\Tr{T_\text{r}} 
\def\Tm{T_\text{m}} 
\def\Trt{T_{\text{r},t}} 
\def\Tat{T_{\text{a},t}} 
\def\uphys{u^\text{phys}}

\def\loss{\mathcal{L}}
\def\lossreg{\loss_{\text{reg}}}
\def\lossphy{\loss_{\text{phy}}}

\begin{document}

\thinmuskip=1\thinmuskip
\medmuskip=1\medmuskip
\thickmuskip=1\thickmuskip

\title[mode = title]{Demand response for residential building heating: Effective Monte Carlo Tree Search control based on physics-informed neural networks}
\shorttitle{DR for residential building heating: Effective MCTS control based on PiNN}

\author[1]{Fabio Pavirani}[orcid=0009-0005-7904-099X]
\ead{fabio.pavirani@ugent.be}
\credit{Conceptualization, Methodology, Software, Formal analysis, Investigation, Visualization, Writing - Original Draft}

\author[1]{Gargya Gokhale}[orcid=0000-0002-1451-397X]
\credit{Software, Conceptualization, Writing - Review \& Editing}

\author[1, 2]{Bert Claessens}[orcid=0009-0006-6116-1483]
\credit{Supervision, Conceptualization, Writing - Review \& Editing}

\author[1]{Chris Develder}[orcid=0000-0003-2707-4176]
\credit{Supervision, Funding acquisition,
Writing - Review \& Editing}

\affiliation[1]{organization={IDLab Ghent university -- imec},
                addressline={Technologiepark Zwijnaarde 126}, 
                postcode={9052}, 
                postcodesep={}, 
                city={Gent},
                country={Belgium}}
\affiliation[2]{organization={Beebop},
                country={Belgium}}
                
\date{June 2023}

\shortauthors{Pavirani et al.}

\bibliographystyle{cas-model2-names}


\graphicspath{{pictures}}

\makeatletter\def\Hy@Warning#1{}\makeatother
\maketitle
\begin{abstract}
To reduce global carbon emissions and limit climate change, controlling energy consumption in buildings is an important piece of the puzzle. 
Here, we specifically focus on using a demand response (DR) algorithm to limit the energy consumption of a residential building's heating system while respecting user's thermal comfort.
In that domain, Reinforcement learning (RL) methods have been shown to be quite effective.
One such RL method is Monte Carlo Tree Search (MCTS), which has achieved impressive success in playing board games (go, chess).
A particular advantage of MCTS is that its decision tree structure naturally allows to integrate exogenous constraints (e.g., by trimming branches that violate them), while conventional RL solutions need more elaborate techniques (e.g., indirectly by adding penalties in the cost/reward function, or through a backup controller that corrects constraint-violating actions).
The main aim of this paper is to study the adoption of MCTS for building control, since this (to the best of our knowledge) has remained largely unexplored.
A specific property of MCTS is that it needs a simulator component that can predict subsequent system states, based on actions taken.
A straightforward data-driven solution is to use black-box neural networks (NNs).
We will however extend a Physics-informed Neural Network (PiNN) model to deliver multi-timestep predictions, and show the benefit it offers in terms of lower prediction errors ($-$32\% MAE) as well as better MCTS performance ($-$4\% energy cost, $+$7\%  thermal comfort) compared to a black-box NN.
A second contribution will be to extend a vanilla MCTS version to adopt the ideas applied in AlphaZero (i.e.,
using learned prior and value functions and an action selection heuristic) 
to obtain lower computational costs while maintaining control performance.

\end{abstract}

\section{Introduction}
\label{sec:intro}

The urgent need to address climate change implies an increased pressure to limit energy usage and mitigate carbon emissions.
In the power grid, this has spurred a significant shift towards ``greener'' energy sources, particularly renewable energy sources (RES).
Yet, since such RES introduce a larger degree of supply uncertainty, maintaining the power grid's balance requires controlling the demand, e.g., through demand response (DR).
Of that demand, a substantial fraction constitutes residential building heating: e.g., residential consumption in the EU amounted to 26.1\%, of which 63.6\% is represented by heating~~\cite{eurostat2020}.
Hence our focus on residential heating system control in this paper.

To realize such residential heating control,
two main strategies have been adopted~\cite{stoffel2023evaluation}: either
\begin{enumerate*}[(i)]
\item \label{it:mpc} model-based controllers, or
\item \label{it:rl} purely data-driven controllers.
\end{enumerate*}
For~\ref{it:mpc}, Model Predictive Control (MPC) is typically used, 
defining a mathematical optimization problem that relies on an explicit model of the building's thermal dynamics.
Solving the problem then delivers (an approximation of) the optimal control decision.
Such model-based techniques, although very efficient, are highly dependent on the model used, 
which typically cannot generalize over different houses~\cite{wang2020reinforcement}.
In terms of~\ref{it:rl}, one of the most used data-driven techniques is Reinforcement Learning (RL)
, which learn from
directly interacting with the environment (\eg the building).
Yet, often a large amount of such interactions (\ie data) is needed to learn a good policy taking (near-)optimal control actions, which can be hard/costly to obtain.
To alleviate this, a simulator (rather than the actual building environment) can be used, but as in MPC, explicitly defining such building models is hampered by the reliance on expert knowledge and lack of generalization across buildings.

Despite their promising results in producing effective control policies, several challenges regarding RL algorithms in the Building Energy Management (BEM) domain are yet to be solved, such as their lack of interpretability and safety~\cite{nagy2023ten}.
Indeed, RL techniques typically rely on data-driven function approximators, such as Neural Networks~\cite{sutton2018reinforcement}, to provide their control actions, resulting in a `black-box' decision-making structure.
The black-box nature implies that the resulting actions cannot be explained/motivated, which triggers reluctance and distrust among users and manufacturers in using such technology~\cite{nagy2023ten}.
Furthermore, since RL agents learn from historical data only, they may take poor decisions particularly in cases they have not encountered before: their actions are not based on any notion of, \eg physical laws governing the system behavior and thus cannot be interpreted~\cite{di2022physically,szegedy2013intriguing}.
This requires backup controllers to override any RL agent actions that would be hazardous~\cite{nagy2023ten}.
In general, it is hard to impose constraints underlying such backup controllers directly into learning an RL policy.

To address these challenges of 
\begin{enumerate*}[(i)]
    \item interpretability, and 
    \item incorporating constraints,
\end{enumerate*}
Monte Carlo Tree Search (MCTS) algorithms are an attractive solution.
Indeed, they are based on estimating the ``value'' of actions (learned from multiple possible rollouts of action sequences), and their tree structure naturally allows to include constraints (e.g., by accordingly pruning the allowed actions from the tree).
Such MCTS solutions relatively recently have achieved significant success in long-standing challenges of board games~\cite{schrittwieser2020mastering}.
However, 
MCTS has remained under-explored in building energy management systems. 
Thus, our paper aims to set baseline performance benchmarks of MCTS solutions in this domain, particularly for heating systems.

We developed an MCTS algorithm that optimizes energy cost while following user-defined constraints.
To simulate the environment dynamics in the MCTS rollouts, we need a model that captures the thermal dynamics of a building.
Starting from a basic data-driven black-box neural network (NN) to encode the system state, we further propose a physics-informed 
Neural Network (PiNN) forecaster.
For the latter, we extend the PhysNet presented in~\cite{gokhale2022physics} to a multi-step prediction model. 
Furthermore, we enhance the standard MCTS structure by incorporating an additional Neural Network (NN) to guide tree construction, similar to the approach proposed in AlphaZero~\cite{silver2016mastering}.
This addition enables the tree search to achieve better actions with lower computational cost, compared to its vanilla version.
Our contributions in studying the effectiveness of MCTS to control a residential heat pump to maximize thermal comfort and minimize energy costs thus can be summarized as follows:
\begin{enumerate}[(1)]
    \item \label{it:mcts-baseline} 
    For modeling the system state transitions in our MCTS solution, we extend the PhysNet model of \citet{gokhale2022physics} to a multi-step forecaster~(\cref{sec:method-pinn}) and show that it offers clear performance benefits, not only in reducing state forecasting errors~($-$32\% MAE; see \cref{sec:results_forecasting}), but also in the eventual MCTS control policy performance~($-$4\% energy cost, $+$7\%  thermal comfort; see \cref{sec:results_mcts});
    \item \label{it:alphazero} Next to the vanilla MCTS solution~(\cref{sec:method-mcts}) that adopts~\ref{it:mcts-baseline}, we propose an AlphaZero-inspired extension that provides non-uniform action priors and an adjusted action selection function. We show that this AlphaZero-MCTS reduces computational cost significantly~(half the number of simulations required to converge; see \cref{sec:results_alphazero}).
\end{enumerate}
The main innovation of this work lies in the combination of \ref{it:mcts-baseline}--\ref{it:alphazero}, by incorporating a physics-informed model into a neural-network-enhanced MCTS solution.
To the best of our knowledge, this is the first such study of MCTS in the demand response domain (in our case for heat pump control), making this work a novel contribution to the field.
\section{Related Work} 
\label{sec:literature_review}
Below we summarize key works that are relevant to our contributions stated above.
Since MCTS solutions require a system model to rollout possible action sequences, we first outline recent common approaches to thermal modeling of buildings (in the context of controlling them) in \cref{sec:literature-modeling}, providing the context for contribution~\ref{it:mcts-baseline}.
Subsequently, in \cref{sec:literature-DR} we list the key control paradigms for building energy management, and thus position MCTS against relevant alternatives including MPC and common RL solutions, to frame our contribution~\ref{it:alphazero}.

\subsection{Thermal dynamics modeling of buildings}
\label{sec:literature-modeling}
Creating an accurate and scalable model that describes the thermal dynamics of a building is crucial to obtain a sequential controller. 
These models can be divided into three main categories: white-box models, black-box models, and hybrid models~\cite{boodi2018intelligent, afroz2018modeling, homod2013review, foucquier2013state, li2014review, deb2017review, bourdeau2019modeling, ali2021review}.

\emph{White-box models} (or first principle models) are typically physics-based methods that model the system using ordinary differential equations~\cite{wetter2014modelica,energyplus}. 
Despite their accuracy, the main drawback of white-box models is the complexity of the physical model developed~\cite{afroz2018modeling}, which limits large-scale adoption (cf.\ a building-specific model needs to be constructed). 

Rather than explicitly modeling the system behavior grounded in fundamental mathematical equations (thus requiring expert knowledge), \emph{black-box methods} purely learn from observations, and thus construct a model of the thermal dynamics directly from historical data, which in principle eases their wide-scale adoption.  
Today, black-box models are typically based on (deep) neural networks (NN).
To properly generalize over a large range of conditions, such NNs need a large and complete dataset, which often is hard to obtain~\cite{di2022physically}.
Furthermore, their black-box nature inherently makes NN models uninterpretable, and they may still take surprising decisions, implying generalization issues~\cite{szegedy2013intriguing}.

\emph{Hybrid models} (or gray-box models) aim to combine the best of both black-box and white-box models, trying to solve their respective limitations while exploiting their benefits. 
The idea is to use generic and usually manageable physical equations together with data-based training to obtain a simple yet accurate model of the building~\cite{afroz2018modeling, afram2015gray, matthiss2023thermal}. 
Because of their simplicity, typically linear RC models are adopted~\cite{bacher2011identifying, vrettos2016experimental}. 

One particular subfamily of hybrid models is based on \emph{Physics-informed Neural Networks (PiNN)}~\cite{raissi2019physics}, 
where prior physical knowledge is incorporated into a data-driven NN prediction model.
The latter is typically achieved by using (possibly simplified) physical laws phrased as partial differential equations, which are then used to regularize NN predictions, \eg by penalizing NN predictions that deviate from the behavior as dictated by physical laws.
By ``infusing'' such physical prior knowledge into a NN model, it can provide a higher level of physical consistency, while still allowing to exploit the NN capacity to capture highly non-linear correlations (that would potentially not be reflected in an approximate physical model).
These characteristics are particularly relevant for the problem of optimal control action search, making PiNN techniques a very valuable solution in building energy control.
Besides their better modeling performance, PiNN has shown promising results regarding their sample efficiency, requiring less data for acceptable results compared to pure black-box models~\cite{gokhale2022physics}~. 

Various types of PiNNs have been applied in modeling the thermal dynamics of buildings from a control-oriented point of view, such as~\cite{gokhale2022physics, drgovna2021physics, di2022physically}.
\citet{gokhale2022physics} modified the loss function of an encoder-based NN to guide it toward physical adherence of the latent state. 
\citet{drgovna2021physics} modeled the state, input, disturbance, and output matrices of a classical linear equation using four different NNs and enforced physical constraints on them.
\citet{di2022physically} modeled the NN architecture to enforce physical consistency on its prediction.
Both the models in~\cite{drgovna2021physics,di2022physically} infuse physical knowledge as an arrangement of the NNs to obtain physical consistency of the predictions. Despite their good results, these networks do not directly access to the prior physical knowledge. Differently from those, the PiNN used in \cite{gokhale2022physics} directly infuses the prior knowledge \textit{inside} the NN equations, by adding a physics loss in the training process.
Moreover, the architecture in \cite{gokhale2022physics} can provide a low-dimensional and physically interpretable latent space in addition to their predictions.
The latent space's values are trained to adhere to relevant and hard-to-measure insight of the building, such as the bulk temperature.
These values can be used, for example, to extend the input taken by controllers to improve their performance.
For this reason, we extended the PhysNet of \citet{gokhale2022physics} towards a multi-step predictor to enable accurate MCTS tree rollouts.
We then incorporated the hidden state in the controller's input, enabling it to make more conscious decisions.

\subsection{Demand response control of buildings}
\label{sec:literature-DR}

With demand response, we refer to control strategies that optimize the energy usage in response to external signals, \eg a time-varying electricity price.
We specifically consider building energy management systems (BEMS), for which~\cite{stoffel2023evaluation} provides a quantitative comparison of state-of-the-art solutions.
In terms of algorithms, they can be roughly classified into two main categories, which we discuss below in more detail: Model Predictive Control (MPC), and Reinforcement Learning (RL) approaches. 

\subsubsection{Model Predictive Control (MPC)}
With MPC, an optimization problem is solved (or approximated) for each control action in a receding horizon approach. 
MPC uses a model to estimate future state transitions (depending on actions taken) and thus requires an accurate model to get satisfactory results. 
As outlined in \cref{sec:literature-modeling}, such models can be either white-box, black-box, or gray-box (hybrid), each with their respective benefits and drawbacks. 
All three approaches have been applied successfully in real applications~\cite{drgovna2020all}. 
For a more detailed list and review of specific MPC variants, we refer to~\cite{yao2021state}. 

White-box MPC models have been shown to be promising when a good model is provided.
Despite this, their application needs a non-negligible amount of expert knowledge in 
defining/selecting the model and framing the mathematical optimization problem, making it difficult to generalize to a broad scale of settings (\eg to multiple varying buildings' parameters). 
To reduce the human effort in designing an MPC controller, data-driven MPC algorithms have received increasing attention.
Among these, deep learning MPC methods using neural networks (NN) are the most popular~\cite{afram2017artificial, lee2015optimal, reynolds2018zone, chen2018lighted}. 
Yet, adopting highly non-linear models such as NN-based ones implies a more complex optimization problem, in the sense that it can no longer be formalized as a linear program (as is typically the case in MPC). 
This complicates solving the optimization problem using readily available commercial solvers~\cite{lawrynczuk2014computationally}


\subsubsection{Reinforcement Learning (RL)}
Another family of data-driven algorithms that has shown promising results in the last decade is Reinforcement Learning (RL). 
RL algorithms formalize the system to be controlled as a Markov Decision Process (MDP) and interact with it to learn an adequate policy that maximizes the expected reward\footnote{Alternatively, minimize the expected cost.} obtained from the environment. 
Such RL controllers have been applied successfully to several problems in the building energy management domain~\cite{brandi2020deep, wei2017deep, patyn2018comparing, nagy2018deep, jiang2021building}. 

Compared to MPC solutions, RL algorithms efficiently learn optimal control actions in a data-driven fashion, thus requiring less expert knowledge to scale the technique.
However, multiple challenges arise with RL techniques~\cite[Question~10]{nagy2023ten}.
For example, RL algorithms often require a large amount of data for training to reach acceptable performance. 
Also, most RL algorithms used in the BEM environment are model-free~\cite{wang2020reinforcement}, which 
makes it hard to incorporate explicitly the sequential planning nature across multiple timesteps, as well as specific constraints.
Instead, model-free RL techniques have to implicitly learn underlying constraints from observed data~\cite{nagy2023ten}.

\subsubsection{Monte Carlo Tree Search (MCTS)}

Monte Carlo Tree Search (MCTS) can be a solution to the problems described above, particularly 
\begin{enumerate*}[(i)]
    \item the dependence of MPC on an explicit system model, and \label{it:mcts-vs-mpc}
    \item the difficulty of RL in dealing with a sequential planning problem as well as specific constraints to be respected. \label{it:mcts-vs-rl}
\end{enumerate*}
Since MCTS just needs the model to be able to role out possible scenarios by simulating them, it is oblivious to the specific model nature, thus alleviating~\ref{it:mcts-vs-mpc} and allowing more scalable/practical models, including black-box or hybrid (gray-box) models.
Given the tree-based nature of MCTS-based planning, it can manage future exogenous constraints naturally by trimming off actions that would lead to violating them (which RL could struggle with, as~\ref{it:mcts-vs-rl} suggests).
This makes MCTS a promising candidate for DR problems where different external signals and objectives need to be followed while handling different operational constraints.

MCTS was first introduced in 2006 by \citet{coulom2006efficient} and gained its popularity thanks to the UCT algorithm~\cite{kocsis2006bandit}.
MCTS has already been applied in various demand response 
solutions for balancing the electrical grid~\cite{wijaya2013effective, galvan2014autonomous, golpayegani2015collaborative, golpayegani2016multi}. 
Specifically for our case of residential demand response on heating systems, the most relevant MCTS work is that of \citet{kiljander2021intelligent}.
They modeled the household thermal dynamics with a Feed Forward Neural Network (FFNN) and used it as a simulator in conjunction with a standard MCTS algorithm to generate planning decisions.
The algorithm was then evaluated in a real household, demonstrating the applicability of the technique in real-world scenarios.
Compared to Kiljander et al., we use a similar conceptual approach using a NN-based simulator in an MCTS framework.
Two notable differences pertain to our contributions listed in \cref{sec:intro}.
First, rather than a black-box NN, we adopt a gray-box, physics-informed NN (PiNN) to model the household, thus improving the physical consistency of the simulator.
Second, inspired by the recent success of~\citet{silver2016mastering,silver2017mastering, silver2018general}, we add a DL layer to the MCTS algorithm to estimate action values, making it more computationally efficient. 
Our problem formulation and PiNN/MCTS methodology is presented in detail in the next section.



\begin{nomenclature}
\begin{deflist}[(A,B,C,D)] 
\defitem{$t$}\defterm{Time-step index}
\defitem{$k$}\defterm{Depth of the node index}
\defitem{$\Delta_t$}\defterm{Length of a time-step}
\defitem{$(X,U,f,\rho)$}\defterm{Stochastic POMDP}
\defitem{$(\tilde{X},\tilde{U},\tilde{f},\tilde{\rho})$}\defterm{Deterministic FOMDP}
\defitem{$x$}\defterm{Full system state of the MDP}
\defitem{$\xbuild\in\Xbuild$}\defterm{Observable building-related state}
\defitem{$\xext\in\Xext$}\defterm{Observable exogenous state}
\defitem{$\xhid\in\Xhid$}\defterm{Non-observable state}
\defitem{$z$}\defterm{PiNN latent state}
\defitem{$\tau$}\defterm{Time of the day}
\defitem{$T_\text{r}$}\defterm{Room temperature}
\defitem{$T_\text{m}$}\defterm{Bulk temperature of the building}
\defitem{$T_{\text{r\_set}}$}\defterm{Desired room temperature}
\defitem{$u$}\defterm{Action signal of the heat pump}
\defitem{$\uphys$}\defterm{Energy consumed by the heat pump}
\defitem{$T_{\text{a}}$}\defterm{Outside temperature}
\defitem{$\lambda$}\defterm{Electricity price}
\defitem{$N(\tilde{x})$}\defterm{Number of tree visit of state $\tilde{x}$}
\defitem{$N(\tilde{x},\tilde{u})$}\defterm{Occurance of action $\tilde{u}$ from $\tilde{x}$ in the tree}
\defitem{$P(\tilde{x},\tilde{u})$}\defterm{Prior probability in the tree traversals}
\defitem{$h$}\defterm{Prediction horizon}
\defitem{$d$}\defterm{Maximum depth of the tree}
\defitem{$l$}\defterm{Leaf node's depth in a tree traversal}
\defitem{$c_1,c_2$}\defterm{Parameters to tune the reward function}
\defitem{$\alpha$}\defterm{Parameter to balance tree exploration}
\defitem{$\Delta_T^{\pm}$}\defterm{Temperature flexibility bounds}
\end{deflist}
\end{nomenclature}

\section{Methodology} 
\label{sec:methods}

\subsection{Markov Decision Process (MDP)}
\label{sec:method-mdp}
We model the sequential control problem of heating a building as a Markov Decision Process (MDP)~\cite{busoniu2017reinforcement}.
An MDP is a mathematical structure described by the tuple $(\mathrm{X}, \mathrm{U}, f, \rho)$ where
$\mathrm{X}$ is the state space,
$\mathrm{U}$ is the action space, $f:\mathrm{X} \times\mathrm{U}\times\mathrm{X}\rightarrow[0, 1]$ is the state transition probability function,
and ${\rho: \mathrm{X} \times \mathrm{U} \times \mathrm{X} \rightarrow \mathbb{R}}$ is the reward function. 
The transition probability function $f(x, u, x')$ gives the probability of transitioning from state $x$ to a new state $x'$ after applying action $u$, whereas the reward $\rho(x, u, x')$ values how ``good'' that action $u$ was.
Through this mathematical structure, we can model the control problem as an optimization problem to find the control policy $\pi: \mathrm{X} \rightarrow \mathrm{U}$ that gives the action $\pi(x)$ to apply in system state $x$ to maximize the expected reward obtained from the environment. 
Formally, given a policy $\pi$, we consider the definition of its Q-function~\cite{busoniu2017reinforcement}, which assesses the long-time reward resulting from applying action $u$ in state $x$:
\begin{equation}
    Q^\pi(x, u) \doteq \mathbb{E}_{x' \sim f(x, u, \cdot)}\left[\rho(x, u, x') + \gamma R^{\pi}(x') \right],
\end{equation}
where:
\begin{equation}
    R^{\pi}(x_0) \doteq \lim_{T\rightarrow +\infty}\mathbb{E}_{x_{t+1} \sim f(x, u, \cdot)}\left[{\sum_{t=0}^{T}{\gamma^t\rho(x_t, \pi(x_t), x_{t+1})}}\right].
\end{equation}
We then consider the definition of the optimal Q-function~\cite{busoniu2017reinforcement}, that is: 
$Q^*(x, u) \doteq \max_\pi{Q^\pi(x, u)}$. 
The optimization problem is then to find an optimal policy $\pi^*$~\cite{busoniu2017reinforcement} such that:
\begin{equation}
    \pi^*(x) \in \arg \max_u{Q^*(x, u)} \; ; \; \forall x \in \mathrm{X} \; .
\end{equation}

MDPs can be classified into two distinct categories: Fully Observable Markov Decision Process (FOMDP) and Partially Observable Markov Decision Process (POMDP).
With FOMDP, the states of the system $x\in\mathrm{X}$ are fully observable by the policy $\pi$.
Conversely, in POMDP only a 
subset of the state variables is observable by the policy, thus we have $\pi:\Xobs\rightarrow\U$, with $\dim(\Xobs) < \dim(\X)$.

The control problem of heating a building is generally represented as a stochastic POMDP, with the objective to minimize the energy cost 
while staying close to the desired temperature set by the users $T_{\text{r\_set}}$. 
We use $t$ to note the time-step index of each system/control variable.
We partition the full system state vector $x_t$ in a building-related one $\xbuild_t$, exogenous variables $\xext_t$ (\eg including current weather conditions at $t$), and unobservable components $\xhid_t$ which remain hidden from our control agent (\eg internal heat gains of the building).
Thus, the full state at timestep $t$ is $x_t = (\xbuild_t, \xext_t, \xhid_t)$ whereas the observable part is limited to $\xobs_t = (\xbuild_t, \xext_t)$.
 
More specifically, we define the observable building state at a time-step $t$ as: ${\xbuild_t=(\Trt, \uphys_{t-1}) \in \Xbuild}$, where $\Trt$ is the room temperature of the building we wish to control and $\uphys_{t-1}$ is the energy consumed by the heating system in the previous time-step. 
The exogenous influences are defined as $\xext_t = (\tau_t, \Tat) \in \Xext$ where $\tau_t \in [0, 24[$ indicates the hour of the day and $\Tat$ is the outdoor temperature.
The action $u \in \mathrm{U}$ is a continuous single value indicating the control signal for the heating component. Higher values of $u_t$ indicate more heating required for timestep $t$, subsequently increasing the energy consumption $\uphys_t$.  
The reward function is defined as:
\begin{flalign}
\label{eq:reward_funct}
    \rho(x_t, u_t) \doteq & \underbrace{-\,\uphys_{t} \, \lambda_t}_\text{Cost optimization}
    && \\ 
    & \underbrace{-\,(\,T_{\text{r\_set}, t} - T_{\text{r}, t+1}\,)^+\, c_1 - (\,T_{\text{r}, t+1} - T_{\text{r\_set}, t}\,)^+\, c_2}_\text{Thermal Comfort optimization} \; , \nonumber
\end{flalign}
where $\lambda_t$ is the energy price at time-step $t$, $T_{\text{r\_set},t}$ is the desired temperature set by the users and $c_1, c_2$ are hyperparameters used to balance the energy cost objective with the user constraints.
In our experiments, the user thermal comfort optimization will use asymmetrical settings ($c_1 > c_2$), because we want to avoid excessive penalization for pre-heating the room.

Ultimately, we will formalize the MDP as a deterministic FOMDP, which will approximate the stochastic POMDP which captures the real building behavior.
That FOMDP is further detailed in \cref{sec:method-control_oriented_modeling}.
As stated before, to model the transition function $f$ in the MDP, which essentially defines the thermal dynamics of the building, we will make use of a PiNN that predicts state transitions, which we describe next.

\subsection{Physics-informed Neural Network (PiNN)}
\label{sec:method-pinn}
\begin{figure}
    \centering
    \includegraphics[width=0.5\textwidth]{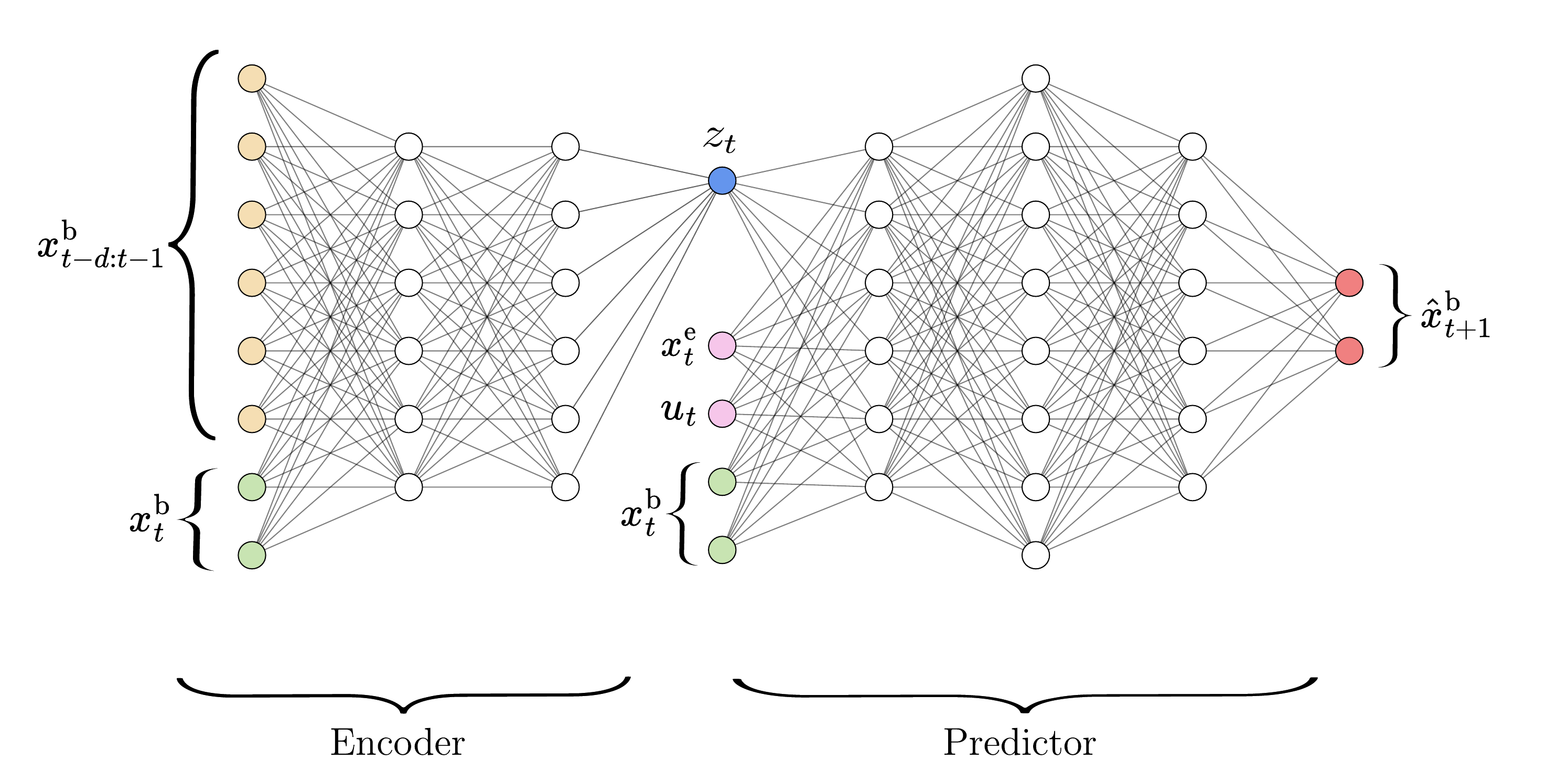}
    \caption{Architecture of the PiNN. Yellow nodes represent past observations (until $t-1$), green nodes are the current observations at timestep $t$, the pink nodes represent the current exogenous state and action, the blue node is the current hidden state, and the red nodes are the predicted observations for the next timestep $t+1$.}
    \label{fig:model_structure}
\end{figure}

We use a PiNN architecture to forecast the next building states of a system $\xbuild \in \Xbuild$ by inserting physical equations into the learning loss function of the NN.
We follow the PhysNet structure proposed by~\citet{gokhale2022physics}, using an encoder-based neural network.
PhysNet is specifically built for projecting a time series spanning the $d \in \mathbb{N}$ most recent building states ${\xbuild_{t-d:t} \doteq (\xbuild_{t-d}, \dots, \xbuild_{t})}$ into a compact hidden state $z_t$, training the NN such that these latent values adhere to certain physical laws.
In particular, $z_t$ will be trained to represent the building mass temperature in a simple RC model of the building's thermodynamic behavior.
A separate predictor NN component will then use that $z_t$, together with observable states (comprising building variables $\xbuild_t$ and exogenous variables $\xext_t$) to predict state transitions, \ie the next building state $\xbuild_{t+1}$, given the current action $u_t \in \mathrm{U}$.
The overall encoder-predictor structure of the NN architecture is illustrated in \cref{fig:model_structure}. 

\begin{figure}
    \centering
    \includegraphics[width=0.5\textwidth]{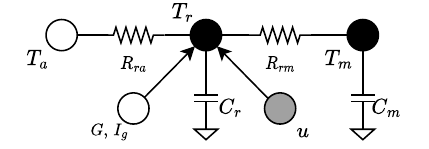}
    \caption{RC model used in the encoder loss~\cite{vrettos2016experimental}. Black nodes \tikzcircle[black, fill=black]{2.5pt} represent building-related state variables, white nodes \tikzcircle[black, fill=white]{2.5pt} the exogenous state variables, and the gray node \tikzcircle[black, fill=black!50]{2.5pt} is the control variable.}
    \label{fig:2R2C_model}
\end{figure}

We expanded the model \cite{gokhale2022physics} to obtain a multi-time-step forecast by deploying a sequential, autoregressive, multi-loss training loop with windowed inference. 
We considered the 2R2C model~\cite{vrettos2016experimental} shown in \cref{fig:2R2C_model} to infuse physical knowledge into the encoder training.
The physical equations obtained are then:
\begin{align}
    \begin{bmatrix}
     \dot{T_\text{r}} \\
     \dot{T_\text{m}}
   \end{bmatrix}
   =
   \begin{bmatrix}
         -\left(\frac{1}{C_\text{r} R_{\text{ra}}} + \frac{1}{C_\text{r} R_{\text{rm}}}\right) & \frac{1}{C_\text{r} R_{\text{rm}}} \\[7pt]
     \frac{1}{C_\text{m} R_{\text{rm}}} & -\frac{1}{C_\text{m} R_{\text{rm}}}
   \end{bmatrix}
   \cdot
   \begin{bmatrix}
        T_\text{r} \\
        T_\text{m}
   \end{bmatrix}
   + \nonumber
   \\
   +
   \begin{bmatrix}
        \frac{c_\text{p}T_\text{s}}{C_\text{r}} \\[5pt]
        0
   \end{bmatrix}
   \cdot
   \dot{u}
   +
   \begin{bmatrix}
        \frac{1}{C_\text{r} R_{\text{ra}}} & \frac{\gamma}{C_\text{r}} & \frac{1}{C_\text{r}} \\[5pt]
        0 & 0 & 0
   \end{bmatrix}
   \cdot
   \begin{bmatrix}
        T_\text{a} \\
        G \\
        I_\text{g}
   \end{bmatrix} \; .
   \label{eq:rc-model}
\end{align}
We are particularly interested in the building mass temperature, indicated as $T_{\text{m}}$. 
For a detailed explanation of the RC model and the notations in \eqref{eq:rc-model}, we refer to \cite{vrettos2016experimental}; the $R_\cdot$ and $C_\cdot$ parameters will be fine-tuned during training of the model.

Note that the encoder-predictor model of \cref{fig:model_structure} only predicts the immediate next observable state (at $t+1$).
To extend the predictions up to a horizon of $h\in\mathbb{N}^+$ timesteps ahead, we autoregressively feed it back to the same encoder-predictor model (\eg to obtain $t+2$ predictions, by using inputs covering the time up to $t+1$, etc.). 
To train the PhysNet, two different losses are adopted for the training loop: a regression loss term for the observable state prediction (\ie to minimize the room temperature prediction error), and a physics loss term to steer the latent state representation (\ie to have $z_t$ as close as possible to the building thermal mass temperature of the RC model):
\begin{equation}
    \loss = \lossreg + \lossphy \;.
\end{equation}
More specifically, the regression loss is an L2 loss between the predicted observable states and the measured ones: 
\begin{equation}
    \lossreg \doteq \sum_{t = 1}^{N}{\left(\xhatbuild_{t+1:t+h} - \xbuild_{t+1:t+h}\right)}^2 \; .
\end{equation}
The physics loss $\lossphy$ is an L2 loss that compares the predicted mass temperatures vector with a target vector $\mathbf{T}_\text{m}^t$ obtained by applying \cref{eq:rc-model}. 
To better understand the training process of the encoder, we introduce the following notations:
\begin{itemize}
    \item We define $\hat{\mathbf{T}}_\text{m}^{t} \in \mathbb{R}^h$ as the vector containing the $h$ autoregressive predictions of the mass temperatures when starting the prediction from the time-step $t$ (\ie the hidden states $z_{t:t+h}$);
    \item We define $\mathbf{T}_\text{r}^{t} \in \mathbb{R}^h$ as the vector containing the $h$ actual measured room temperatures, starting from the time-step $t$.
\end{itemize}
The target value for the encoder training when starting from the time-step $t$ is then defined by applying \cref{eq:rc-model} as: 
\begin{equation}
\label{eq:encoder_training}
    \mathbf{T}^t_\text{m} = \mathbf{\hat{T}}_\text{m}^{t-1} + \frac{\Delta_t}{C_\text{m} R_{\text{rm}}} \left( \mathbf{T}_\text{r}^{t-1} - \mathbf{\hat{T}}_\text{m}^{t-1} \right) \; ,
\end{equation}
where $\Delta_t$ is the duration of a time-step. 
The loss function to be minimized will then be:
\begin{equation}
\label{eq:physical_loss}
    {\mathcal{L}_{\text{phys}} \doteq \sum_{t = 1}^{N}{\left(\mathbf{\hat{T}}_\text{m}^t - \mathbf{T}_\text{m}^{t}\right)}^2} \; .
\end{equation}
To study the impact of the PiNN component, we will also perform a black-box ablation that omits $\mathcal{L}_{\text{phys}}$ (see \cref{sec:results_forecasting}).
More details about the neural networks hyperparameters are in \cref{appendix:hyperparameters}.

\subsection{Control-oriented modeling}
\label{sec:method-control_oriented_modeling}
To approach the control power described in \cref{sec:method-mdp}, we will adopt an MCTS technique that requires a deterministic MDP with a discrete action space.
In particular, our MCTS agent will rely on a deterministic, fully observable MDP (FOMDP) that approximates the stochastic partially observable MDP (POMDP) that we introduced in \cref{sec:method-mdp}.
We consider a FOMDP with deterministic dynamics described by the PhysNet introduced in \cref{sec:method-pinn}, \ie the exogenous information $\xext_t$ gets forecasted for a fixed horizon $h$,
thus enabling MCTS to deterministically simulate a state trajectory based on chosen (heating) actions.

To differentiate it from the real household POMDP, we indicate this approximated environment with a different notation: $(\tilde{\mathrm{X}}, \tilde{\mathrm{U}}, \tilde{f}, \tilde{\rho})$.
The state at a certain time-step $t$ is a continuous vector defined as:
\begin{equation}
    \tilde{\mathbf{x}}_t \doteq (\tau_t,\; T_{\text{r},t},\; T_{\text{m},t}) \in \tilde{\mathrm{X}}
\end{equation}
where: $\tau_t \in [0, 24[$ is the time of the day, $T_{\text{r},t} \in \mathbb{R}$ is the room temperature, and $T_{\text{m},t} \in \mathbb{R}$ is the (estimated) mass temperature of the building (as predicted by the PiNN).
The action space is discrete and single-dimensional $\tilde{\mathrm{U}} \subset \mathrm{U}$, representing the heating system control action.
The transition function $\tilde{f}$ is a trained PiNN as described in \cref{sec:method-pinn}. Last, the reward function $\tilde{\rho}$ is defined, similarly to the original one introduced in \cref{eq:reward_funct}, as:
\begin{flalign}
    \tilde{\rho}(\tilde{x}_t, \tilde{u}_t) \doteq & \underbrace{-\,\uphys_{t} \, \lambda_t}_\text{Cost optimization}
    && \nonumber \\ 
    & \underbrace{-\,(\,T_{\text{r\_set}, t} - T_{\text{r}, t+1}\,)^+\, c_1 - (\,T_{\text{r}, t+1} - T_{\text{r\_set}, t}\,)^+\, c_2}_\text{Thermal Comfort optimization} \label{eq:reward_funct2}
\end{flalign}
This structure provides the capability to simulate future states --- which approximate the real dynamics through using a PiNN --- enabling 
the use of MCTS.

\subsection{Monte Carlo Tree Search (MCTS)}
\label{sec:method-mcts}
In MCTS, the possible scenarios that the current state may evolve into are represented in a tree structure, modeling subsequent states of the environment as nodes and the actions governing transitions between them as edges. 
We will consider deterministic environments with the assumption that a generative model of the MDP is available (\ie the environment can be simulated), as explained in \cref{sec:method-control_oriented_modeling}.
As introduced there, 
we will refer to the state space of the environment as $\tilde{\mathrm{X}}$, the action space as $\tilde{\mathrm{U}}$, the (deterministic) transition function as $\tilde{f}: \tilde{\mathrm{X}} \times \tilde{\mathrm{U}} \rightarrow \tilde{\mathrm{X}}$ and the reward function as ${\tilde{\rho}: \tilde{\mathrm{X}} \times \tilde{\mathrm{U}} \rightarrow \mathbb{R}}$. 

The ultimate goal of MCTS (or any tree search algorithm) is to enable to search for the most promising action to take from each system state, \ie the one that allows to reach maximal reward. 
Thus, it will attach a ``value'' to each node --- usually formalized as a Q-value function --- representing how rewarding it is to move to that state, and do so based on simulation results.
The general structure of the MCTS algorithm comprises four sequential phases that are repeated iteratively, until an acceptable solution is obtained (\ie the node ``values'' are deemed to be representative for making good decisions): 
\begin{enumerate}
    \item \textbf{Selection:} select 
    a node (\ie a system state) to further explore actions for.
    \item \textbf{Expansion:} expand the tree by adding new node(s), \ie roll out possible actions from that state.
    \item \label{it:simulation_phase} \textbf{Simulation:} evaluate the node value by performing Monte Carlo simulations. 
    \item \textbf{Backpropagation:} propagate the information acquired back to the root node.
\end{enumerate}
After iterating through these phases (``searching''), the tree is used at inference time to select the actions based on the node values they lead to.
For selection strategies like UCT, it has been proven that the search converges to the optimal solution~\cite{kocsis2006bandit}. 
Recent research has advanced MCTS performance by changing the UCT score formula \cite{rosin2011multi} and adding NNs for value and policy estimation in the tree search~\cite{silver2016mastering, silver2017mastering, silver2018general, schrittwieser2020mastering}.

In our experiments, we will particularly 
adopt improvements as proposed in the work of~\cite{silver2018general, schrittwieser2020mastering}.
Following \cite{silver2018general}, we avoid random rollouts by Monte Carlo simulations, but rather use a trained NN to approximate the Q-values based on more targeted exploration of the tree.
This implies that the Simulation step in the original MCTS algorithm above can be omitted.
Thus, our \emph{Vanilla MCTS} algorithm 
comprises the following steps:
\begin{enumerate}
    \item \textbf{Selection:}
    \label{it:vanilla-mcts-selection}
    Starting from a root node $\tilde{x}^0 \in \tilde{\mathrm{X}}$, the current tree is traversed by selecting the action $\tilde{u}^k \in \tilde{\mathrm{U}}^k, k \in \mathbb{N}$, until a leaf node $\tilde{x}^\ell$ is reached. 
    Note that we will restrict the possible actions $\tilde{\mathrm{U}}^k \subseteq \tilde{\mathrm{U}}$ from a given state $\tilde{u}^k$, based on constraints from our environment (\eg no heating possible if the temperature exceeds the user setpoint beyond a given tolerance).
    The selection of each action in this phase is based on:\footnote{Note that we omitted a portion of the formula used in \cite{schrittwieser2020mastering}. Specifically, we removed $\log\left( \frac{N(\tilde{x}^k) + c+1}{c} \right)$ as it would become null when~${c \gg N(\tilde{x}^k)}$, which is the case when using the values suggested in the original work.}
    \begin{flalign}
    \label{eq:modified_uct_score}
        \tilde{u}^k = \arg \max_{\tilde{u}\in\tilde{\mathrm{U}}^k}{\biggl\{\underbrace{\tilde{Q}(\tilde{x}^k, \tilde{u})}_\text{Value score} +  \underbrace{\alpha \> \frac{\sqrt{N(\tilde{x}^k)}}{1 + N(\tilde{x}^k, \tilde{u})}}_\text{Exploration score}\biggl\}} ; 
        && \\ \nonumber \forall k \in 0, 1,  \dots, \ell-1
    \end{flalign}
    where $N(\tilde{x})$ is the number of times the node $\tilde{x}$ has been reached so far, 
    $N(\tilde{x},\tilde{u})$ is the number of times the action $\tilde{u}$ has been selected, 
    and $\tilde{Q}(\tilde{x},\tilde{u})$ is a state-action value function that estimates the expected future average rewards obtained by executing the action $\tilde{u}$ from state $\tilde{x}$.
    (After this selection, the $N(\cdot)$ values along the followed path will be updated accordingly.)
    The balance between exploration and exploitation can be regulated by the hyperparameter $\alpha > 0$. 
    In our experiments, we set $\alpha = 1$, unless stated otherwise.
    
    \item \textbf{Expansion:}
    \label{it:vanilla-mcts-expansion}
    Upon reaching a leaf node $\tilde{x}^\ell$, if its depth does not exceed a fixed maximum value, it is expanded with new nodes (thus adding to the tree constructed so far), one for each possible action $\tilde{u} \in \tilde{\mathrm{U}}^\ell$. 

    \item \textbf{Backpropagation:}
    \label{it:vanilla-mcts-backprop}
    After the expansion phase, 
    reward information gets backpropagated to the Q-values 
    along the selected trajectory, from the original leaf node $\tilde{x}^\ell$ upwards to the root node, as follows:\footnote{Note that compared to \cite{schrittwieser2020mastering}, we added the $\ell - k$ division of the $G^{k+1}$ term. This leads to normalized Q-values that are contained in $[0,1]$.}
    \begin{align}
    \label{eq:mcts_q_update} 
        \tilde{Q}(\tilde{x}^k, \tilde{u}^k) = \frac{N(\tilde{x}^k, \tilde{u}^k) \tilde{Q}(\tilde{x}^k, \tilde{u}^k) + \frac{G^{k+1}}{(\ell-k)}}{N(\tilde{x}^k, \tilde{u}^k) + 1} ; \\
        \nonumber \forall k = \ell-1, \dots, 0
    \end{align}
    where $\tilde{Q}(\tilde{x}^k, \tilde{x}^k)$ is initialized as $\tilde{\rho}(\tilde{x}^k, \tilde{u}^k)$ and $G^k$ is the discounted accumulated reward defined as:
    \begin{equation}
        \begin{cases}
          G^\ell = \tilde{\rho}(\tilde{x}^{l-1}, \tilde{u}^{\ell-1}) \\
          G^k = \tilde{\rho}(\tilde{x}^{k-1}, \tilde{u}^{k-1}) + \gamma \> G^{k+1} 
          \quad \forall k = \ell-1, \dots, 1
        \end{cases} \nonumber
    \end{equation}
    where $\gamma$ is the discount value. 
    %
\end{enumerate}
The rewards $\tilde{\rho}(\tilde{x}, \tilde{u})$ considered in this algorithm are normalized between $0$ and $1$, ensuring that every $\tilde{Q}$ value in \cref{eq:mcts_q_update} is also contained in $[0, 1]$.
At the end of iterating through these phases, the resulting tree is used to choose the most selected\footnote{Or, equivalently, the one with the highest $\tilde{Q}(\tilde{x}, \tilde{u})$ value.} action $\tilde{u}$ from the current system state $\tilde{x}$ (\ie the root of the tree that was built). 

To further evaluate MCTS techniques in the BEM domain, we also deployed a more advanced version that uses a NN to lead the tree search through more optimal nodes, similarly to what was proposed in AlphaZero~\cite{silver2018general}.
Our resulting \emph{AlphaZero MCTS} algorithm iterates over the same selection/expansion/backpropagation phases, but modifies the Selection phase (step~\ref{it:vanilla-mcts-selection}) by changing \cref{eq:modified_uct_score} to:
\begin{flalign}
\label{eq:alphazero_uct}
    \tilde{u}^k = \arg \max_{\tilde{u}\in\tilde{\mathrm{U}}^k}{\biggl\{\underbrace{\tilde{Q}(\tilde{x}^k, \tilde{u})}_\text{Value score} +  P(\tilde{x}^k, \tilde{u}) \> \underbrace{\alpha \> \frac{\sqrt{N(\tilde{x}^k)}}{1 + N(\tilde{x}^k, \tilde{u})}}_\text{Exploration score}\biggl\}} \, ; 
    && \\ \nonumber \forall k \in 0, 1,  \dots, \ell-1
\end{flalign}
Here $P(\tilde{x}, \tilde{u})$ is a prior probability value associated with taking action $\tilde{u}$ from state $\tilde{x}$.
This probability will be learned by a NN (with $\tilde{x}$ as inputs and a probability distribution over $\tilde{u} \in \tilde{U}$ as output).
This NN is trained offline by using sampled trajectories obtained with the Vanilla MCTS algorithm, considering the FOMDP simulator for both the search and evaluation environment.
Details on the NN structure are listed in \cref{appendix:hyperparameters}.
For balancing exploration/exploitation, when using AlphaZero MCTS, we set $\alpha = 3.5$.\footnote{Note that the higher $\alpha$ in AlphaZero MCTS as opposed to Vanilla MCTS stems from the additional multiplication with $P(\tilde{x},\tilde{u}) \leq 1$.} 

\section{Experiment Setup}
\label{sec:experiment-setup}

\subsection{Building model} 
\label{sec:experiment-building-model}
As a building model representative of real-world conditions, we used the BOPTEST benchmark framework developed by \citet{blum2021building}.
To efficiently integrate BOPTEST with our python-based implementation of the aforementioned PiNN and MCTS components, we used the OpenAI Gym toolkit BOPTEST-Gym \cite{arroyo2021open}.
The specific model we used is their single-zone Hydronic Heat Pump simulation, representing a five-member residential dwelling located in Brussels, Belgium.
The single-zone model considers a $12\times16\,\text{m}^2$ rectangular floor, heated with an air-to-water modulating heat pump of $15\,\text{kW}$ nominal heating capacity.
The action space consists of the modulation signal of the heat pump, $\mathrm{U} \doteq [0,1]$.

Because the MCTS algorithm assumes discrete actions, we consider 5 equally spaced actions: $\tilde{\mathrm{U}} \subset \mathrm{U} \doteq [0, 1]$.
To analyze our MCTS algorithm's potential, we consider two different price scenarios:
\begin{enumerate*}[(i)]
\item one with real-world day-ahead price data from BELPEX of 2019,\footnote{The price data used is provided by BOPTEST \cite{blum2021building} as the ``higly dynamical electricity price'' scenario, see \url{https://ibpsa.github.io/project1-boptest/testcases/ibpsa/testcases_ibpsa_bestest_hydronic_heat_pump/}.} and
\item a synthetic one using a square wave profile.
\end{enumerate*}

\subsection{Constraints}
\label{sec:experiment-constraints}
An advantage of MCTS is that it can easily incorporate dynamic (\ie state-dependent) constraints on the possible actions (\cf our reduced action space $\tilde{\mathrm{U}}^k \subseteq \tilde{\mathrm{U}}$ from a current state $\tilde{x}^k$), 
which amounts to pruning action edges from the tree.
Such backup constraints can be picked with total freedom based on the available state values, allowing for rather complex constraints, beyond simple clipping.
This enables more efficient constraint management compared to classic RL agents, where backup actions are enforced a posteriori (\ie correcting the chosen RL actions if need be), without the RL controller being aware of them.

As an example of such constraints (which could be easily adapted), we will specifically
assume a backup controller that forces the heat pump to activate (respectively turn off) when the room temperature falls below the desired one by $\Delta_\text{T}^-$ (respectively exceeds it by $\Delta_\text{T}^+$).

\subsection{Validation objectives and metrics}
\label{sec:experiment-objectives}
Our experiments discussed in the next section have as primary objectives to evaluate the performance of both 
\begin{enumerate*}[(1)]
    \item the system state forecasting component, \ie our PhysNet extension (from \cref{sec:method-pinn}), and
    \item the MCTS-based controller (from \cref{sec:method-mcts}).
\end{enumerate*}

\begin{table*}[t]
\centering
\captionof{table}{Description of the data used}
    \begin{tabular}{lll}
        \toprule
        Symbol & Description & Value domain \\
        \midrule
        $\tau$ & Time of the day [h] & $\{0, 0.5, 1, \ldots, 23.5\}$ \\
        $T_\text{r}$ & Indoor temperature [\centigrades] & $[15,\> 25]$ \\
        $T_\text{a}$ & Outside temperature measured in a dry bulb [\centigrades] & $[-10, \> 20]$ \\
        $u$ & Heat pump modulating signal for the compressor speed & $[0, \> 1]$ \\
        $\uphys$ & Heat pump electrical power consumed [W] & $[0, \> 4000]$ \\
        $\lambda$ & Belgian day-ahead energy prices as determined by the BELPEX electricity market [€/kWh] & $[-0.4, \> 0.4]$ \\
        \bottomrule
    \end{tabular}
\label{tab:generated data description}
\end{table*}

To evaluate the PhysNet forecasting ability, we used BOPTEST-Gym to generate the relevant data time series at 30\,min resolution by simulating the heat pump usage with a continuous controller (details in \cref{appendix:rulebased_controller}) starting from January 1, 2019.
The parameters that we record the time series for are listed in \cref{tab:generated data description}.
We add cumulative noise to the outside temperature to emulate the prediction error of a weather forecaster tool (see \cref{appendix:cumulative_noise} for more details).
To assess the benefits of infusing physical knowledge into the model, we benchmark it against a black-box ablation that shares the same architecture but omits the physical loss defined in \cref{eq:physical_loss}.
As a performance metric, we use the Mean Absolute Error (MAE) between model predictions and recorded values.

\begin{figure}
    \centering
    \includegraphics[width=0.49\textwidth]{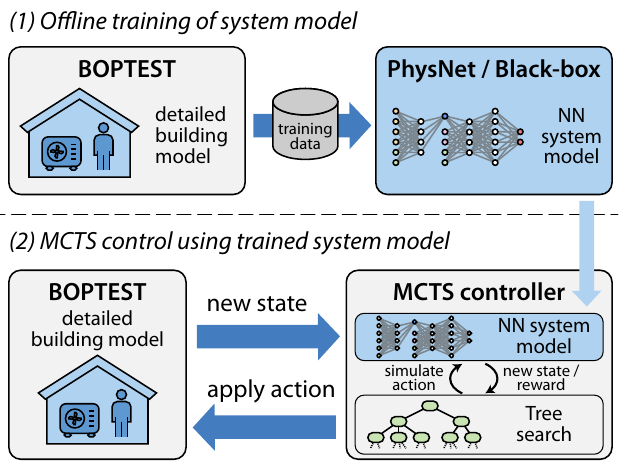} 
    \caption{Overview of our proposed MCTS control strategy. The NN-based system model is first trained with historical data. It is then used in MCTS for simulating action roll-outs, from which the tree search algorithm eventually decides on its optimal action to take, which is then applied to the actual system (in BOPTEST). These steps are repeated as needed.}
    \label{fig:project_structure}
\end{figure}

Besides evaluating the predictive performance of PhysNet, we also aim to evaluate the benefit it brings (compared to a black-box baseline) when adopted in our proposed MCTS controller.
To train the system state transition models (PhysNet and black-box), we simulate the selected building environment with BOPTEST-Gym for 10 days, using a discrete rule-based controller (details in \cref{appendix:rulebased_controller}) to take the heating actions.
%
We then use our trained networks (PhysNet or black-box) for simulating system state transitions in building the trees in the MCTS algorithms described in \cref{sec:method-mcts}, starting from the current state in BOPTEST-Gym.
Based on the tree built, the action to apply is chosen, executed in BOPTEST-Gym, which then provides the actual state at the following decision timepoint.
For that new state, a new MCTS tree is built, and the process repeats.
The PhysNet/black-box system model's NNs are retrained after each day of evaluation, by adding the newly obtained data (with the MCTS controller's actions) to the initial 10-day training set.
We consider a test period of 11 consecutive days, following the 10 initial ones.
\Cref{fig:project_structure} illustrates this setup.

To quantitatively evaluate our controller's performance, we use the average cumulative reward (defined in \cref{eq:reward_funct2}) obtained from the environment over the course of a day: higher rewards indicate better adherence of room temperatures to desired levels, achieved while minimizing costs.
We normalize this reward per timestep to [0,1], resulting in a maximal reward of 48 over a single day, given that we use 30\,min timesteps. We provide full details about the normalization process in \cref{appendix:normalization_rewards}.
To further benchmark our MCTS controllers' results, we also consider a price-agnostic rule-based controller, namely a bang-bang controller that naively heats the room whenever its temperature falls below the desired one (for details see \cref{appendix:rulebased_controller}).

\section{Results}
\label{sec:results}
As explained in \cref{sec:experiment-objectives}, 
our experimental results will subsequently answer the following research questions:
\begin{Q}
    \item Is the PhysNet model more \textit{accurate and data-efficient} 
    than a black-box model? (\cref{sec:results_forecasting})
    \item Is the PhysNet model able to provide \textit{better control actions} when used in a Vanilla MCTS algorithm, compared to a black-box model? (\cref{sec:results_mcts})
    \item What computational benefits arise from a more targeted tree search exploration, \ie using AlphaZero MCTS instead of Vanilla MCTS? (\cref{sec:results_alphazero})
\end{Q}

\subsection{Forecasting results}
\label{sec:results_forecasting}
\begin{figure*}[t]
    \centering
    \includegraphics[width=0.8\textwidth]{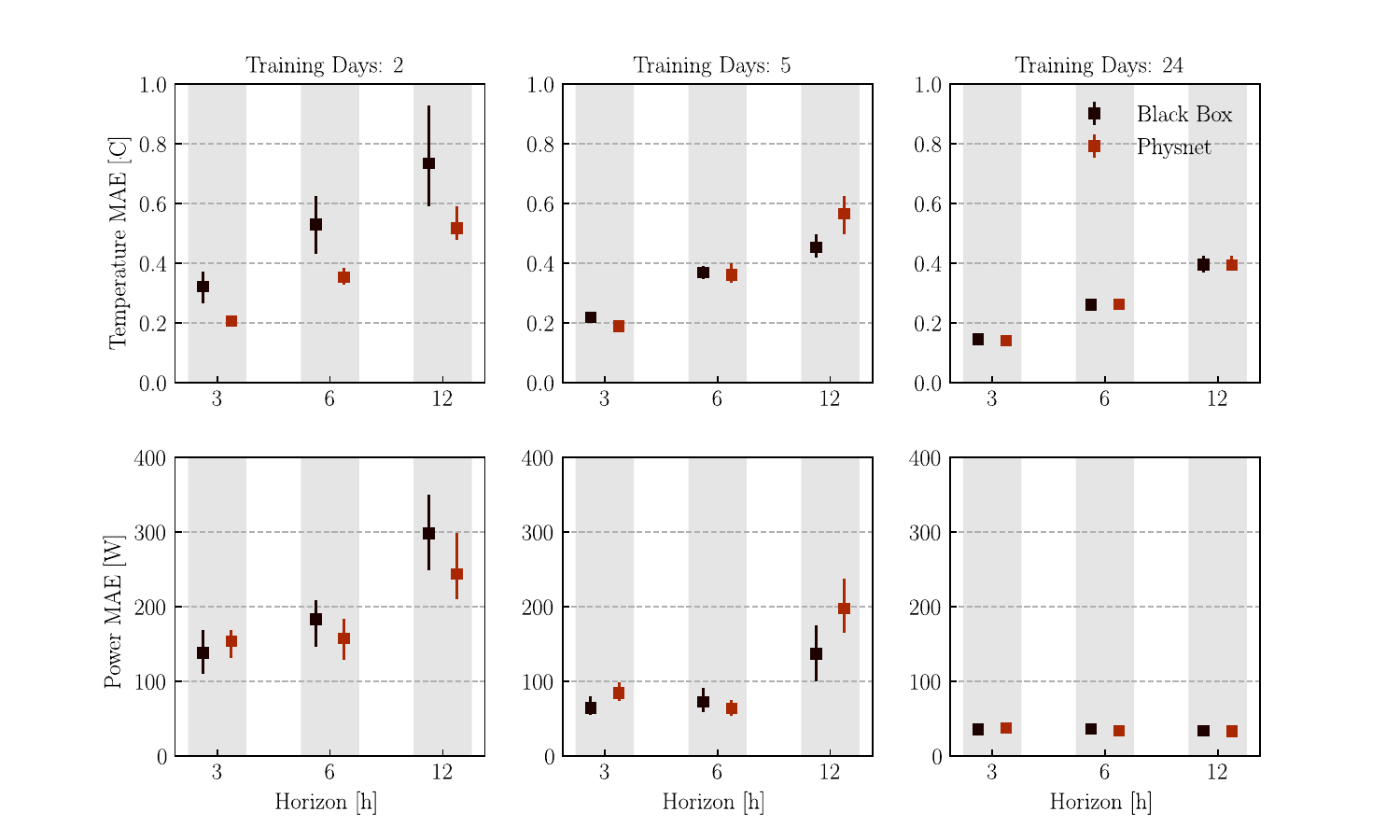}
    \caption{Prediction results of the PhysNet architecture and its black-box counterpart with different training sizes and prediction horizons. PhysNet's temperature predictions outperform the black-box ones when a small dataset (2~days) is considered. The performance of the two architectures converges to the same prediction errors when a bigger training set (24~days) is used.}
    \label{fig:results_forecasting}
\end{figure*}

We first compare the forecasting performance of the PiNN model to its black-box counterpart. 
Training sets of 2, 5, and 24~days were generated with the BOPTEST framework, each for prediction horizons of 3, 6, and 12~hours.
As indicated in \cref{sec:experiment-objectives}, we adopt a continuous controller (see \cref{appendix:rulebased_controller}) to decide on heating actions in these scenarios.
We evaluate each trained model with a test set spanning 6~days.
To allow robust comparison, we average results over 100 runs (varying only the random initialization seed) for each model.
The Mean Absolute Error (MAE) results are shown in \cref{fig:results_forecasting}, where the scatter plot represents the median value and interquartile range for each model combination (architecture + training length + prediction horizon).

We first focus on a small training dataset (\ie the first column of \cref{fig:results_forecasting}, with 2 days of training data).
Regarding the temperature predictions (top row), we note a substantial improvement of the PhysNet compared to its black-box ablation.
For each prediction horizon, PhysNet achieves lower prediction error (with an average reduction of 32\%) as well as a more stable performance (\ie the interquartile bands are smaller) compared to the black-box model.
Looking at the energy consumption predictions (bottom row of graphs in \cref{fig:results_forecasting}), PhysNet still outperforms the black-box model for higher prediction horizons, but with 
larger performance variation (\cf larger interquartile whiskers) and 
less pronounced MAE benefits compared to the black-box model (10\% reduction on average).
Given the PhysNet architecture, these results are in line with our expectations.
The PhysNet addition consists of physics prior knowledge regarding the bulk temperature of the building, which mostly defines the thermal exchange of the room with the structure's mass.
These exchanges mostly influence the room temperature, and not the energy consumed by the heater component.
For this reason, we did expect a more pronounced improvement of the PhysNet in the temperature predictions compared to the energy consumption ones.

When increasing the training data to 5 days, the black-box models achieve a noticeable MAE reduction compared to when a smaller training set is considered.
Conversely, PhysNet predictions do not benefit from a similar improvement, especially on the temperature side, where the MAE for a 5-day training set is on par with the 2-day based model.
Finally, when increasing the training data to 24 days, both models converge to the same results with very high stability in both temperature and energy consumption predictions.
This result outlines that --- when solely considering the forecasting metric (\ie MAE) --- these two models achieve equivalent results when fed with a big enough training dataset.
Nonetheless, with a small amount of data, PhysNet seems to outperform the black-box model, especially with regard to the temperature forecasts.

\begin{figure}
    \centering
    \includegraphics[width=0.45\textwidth]{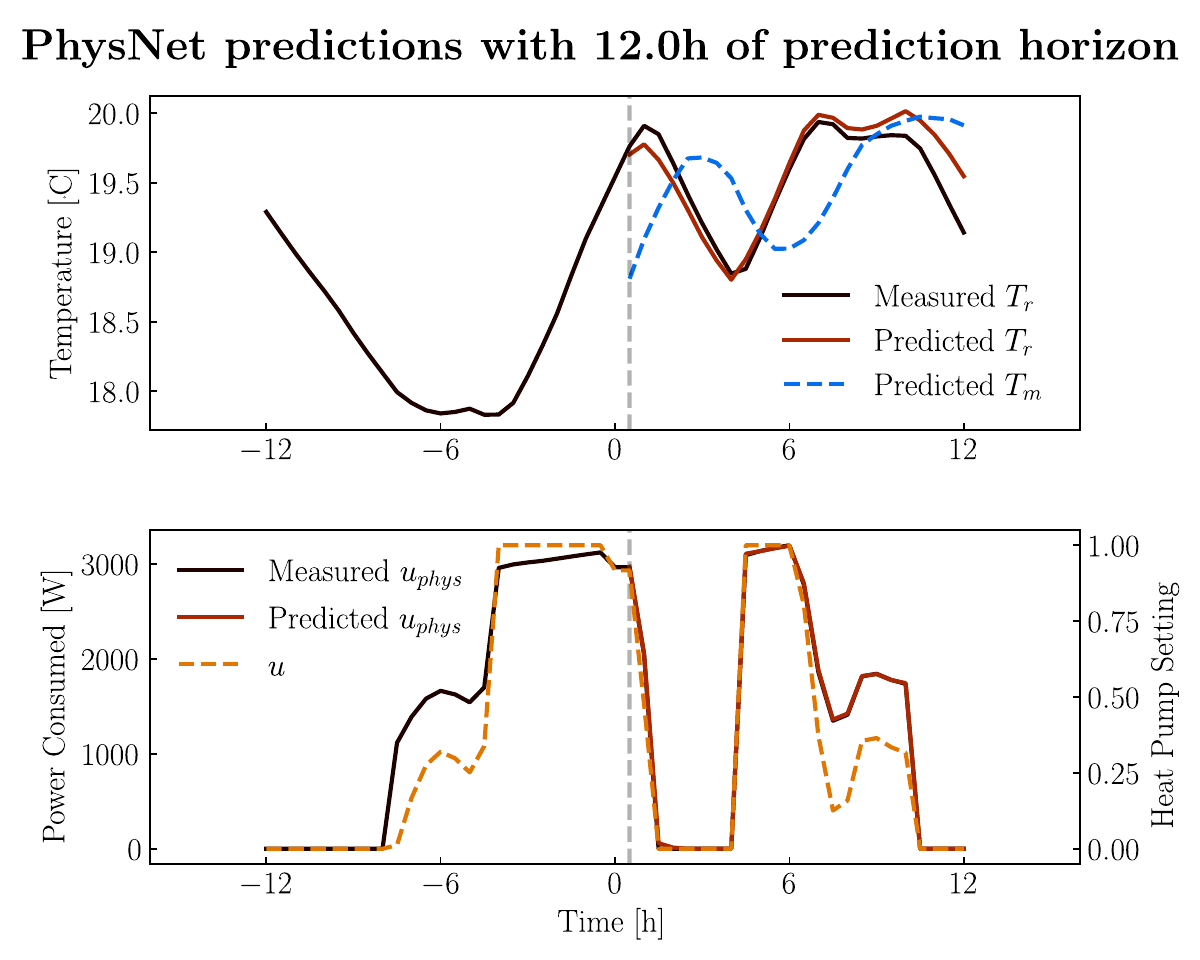}
    \caption{PiNN predictions over a horizon of 12 hours. The temperature graph on top illustrates the expected behavior of the building mass as thermal storage that exchanges heat with the room at a slower pace, \ie $\Tm$ lags behind $\Tr$.}
    \label{fig:2state_prediction}
\end{figure}

Finally, we wanted to validate the mass temperature estimation of PhysNet.
Yet, since building mass temperature cannot be easily measured in practice, we provide qualitative rather than quantitative evaluation.
To verify that the predicted mass temperature $\Tm$ behaves as expected, \cref{fig:2state_prediction} presents the predicted quantities over a horizon of 12\,h.
We observe that the predicted $T_{\text{m}}$ varies with a higher delay compared to $T_\text{r}$, consistent with the expected behavior of a building mass's greater inertia.

Through this first experiment, we established the value of our models, numerically evaluating their performance as forecasters.
We also highlighted the first benefits of the PiNN model compared to the black-box one, namely its higher data efficiency and its hidden state generation that correctly resembles an insightful and hard-to-measure physical value of the building.

\subsection{PiNN control benefits}
\label{sec:results_mcts}
\begin{figure}
    \centering
    \includegraphics[width=0.45\textwidth]{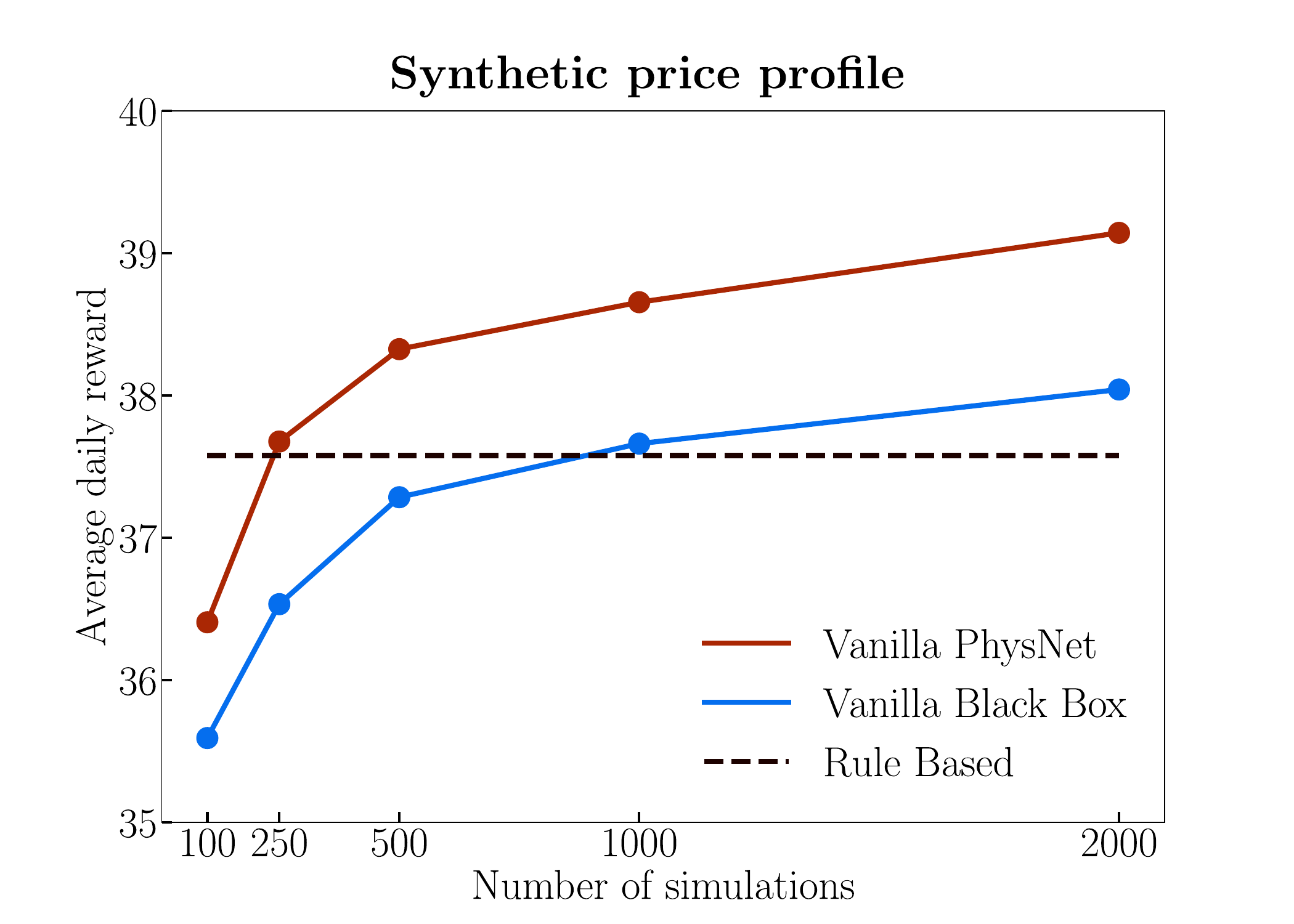}
    \includegraphics[width=0.45\textwidth]{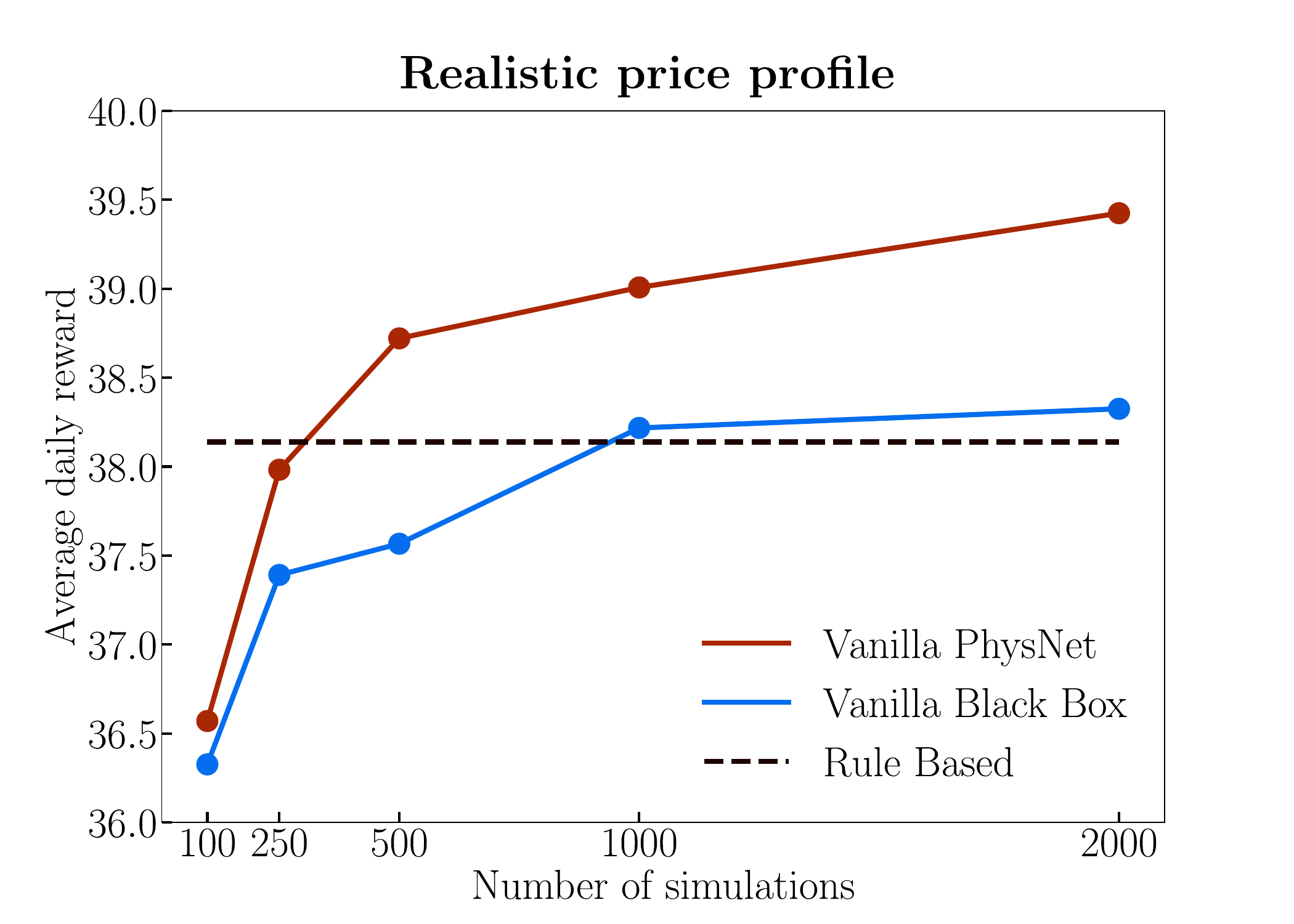}
    \caption{Average daily rewards obtained by the Vanilla MCTS when applied to 11 test days with a PhysNet and a black-box simulator. The performance obtained with the PhysNet model remarkably outperforms the one obtained with the black-box simulator.}
    \label{fig:PiNN_vs_bb_results}
\end{figure}

\begin{figure}
    \centering
    \includegraphics[width=0.45\textwidth]{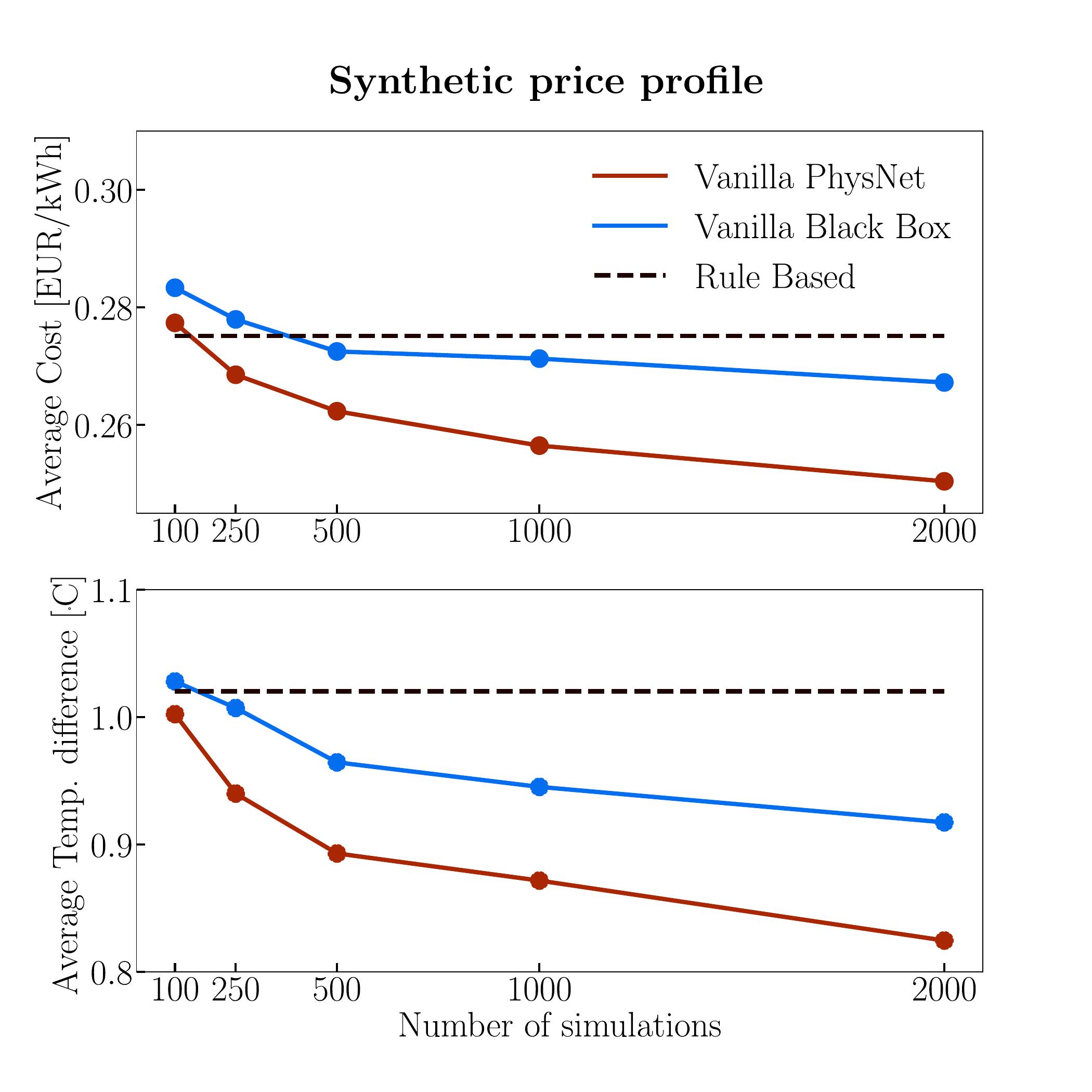}
    \caption{Average cost per kWh for powering the heat pump following the controller's heating actions (first row) and average of the absolute value of the room temperature deviation from the desired one (second row). Conforming with the results shown in \cref{fig:PiNN_vs_bb_results}, the MCTS controllers achieve lower costs and more room temperature adherence to the desired one.}
    \label{fig:cost_temp_plot}
\end{figure}

We now focus on the hypothesized benefit of using a PiNN-based system model within an MCTS controller.
We plug the PhysNet (and its black-box reference model) into our Vanilla MCTS controller, which we evaluate at different levels of computational complexity: we vary the number of times we cycle through the selection/expansion/back\-propagation steps. 
This number of iterations, which affects the eventual MCTS tree depth, is commonly referred to as `number of simulations'.
The resulting controller performance is shown in \cref{fig:PiNN_vs_bb_results} (daily reward) and \cref{fig:cost_temp_plot} (monetary cost and temperature setpoint deviation).

We first focus on the differences between the MCTS application with a PhysNet and a black-box model.
We observe in \cref{fig:PiNN_vs_bb_results} that Physnet enables the algorithm to obtain significantly higher daily rewards (in the range of $+$3\%) compared to the black-box model, and consistently does so when increasing the number of simulations to construct the MCTS tree.
On the latter, we observe that performance improves logarithmically with increasing number of simulations.
This suggests that the main control behavior is learned relatively fast, but further improvements are achieved through deeper tree rollouts. 
Intuitively, it makes sense that expanding that forward-looking horizon very far only brings limited benefit (hence the diminishing benefit in daily reward).

Given that the prediction performance differences observed in \cref{sec:results_forecasting} between black-box and PhysNet models seemed limited, it may be rather surprising that when deployed in an MCTS loop they lead to significantly different results.
We hypothesize that this is due to the higher physical consistency of the PhysNet model, which allows it to better predict the system behavior for control actions that are ``out-of-distribution'' with respect to the training data (as observed in~\citet{di2022physically}).
Indeed, in building the MCTS tree, we need accurate predictions for a wide range of actions, \eg heating when it's already hot, or conversely not heating when the temperature gets too low.
Such scenarios will not be represented in the training set for the system model, and PhysNet can rely on its ``physical knowledge'' to better predict the corresponding state trajectories than a purely data-driven black-box counterpart.

To contextualize the general performance of our MCTS solution, we compare it against the naive rule-based (bang-bang) controller that simply heats whenever the temperature drops below the setpoint.
We note that the black-box controller needs MTCS to use 1,000 simulations to be able to beat bang-bang, whereas the PhysNet-based MCTS already matches it at $\sim$250 simulations and significantly outperforms it for more.
When looking at the actual system parameters of interest to the user (\ie monetary cost and effective room temperature; both reflected jointly in the reward function \cref{eq:reward_funct2}), as plotted in \cref{fig:cost_temp_plot} for the synthetic price profile, we note qualitatively similar performance benefits in comparing PhysNet MCTS controllers to the black-box counterpart (average cost reduction of 4\% and temperature deviation reduction of 7\% for PhysNet), as well as compared to the rule-based baseline.


\begin{figure}
    \centering
    \includegraphics[width=0.45\textwidth]{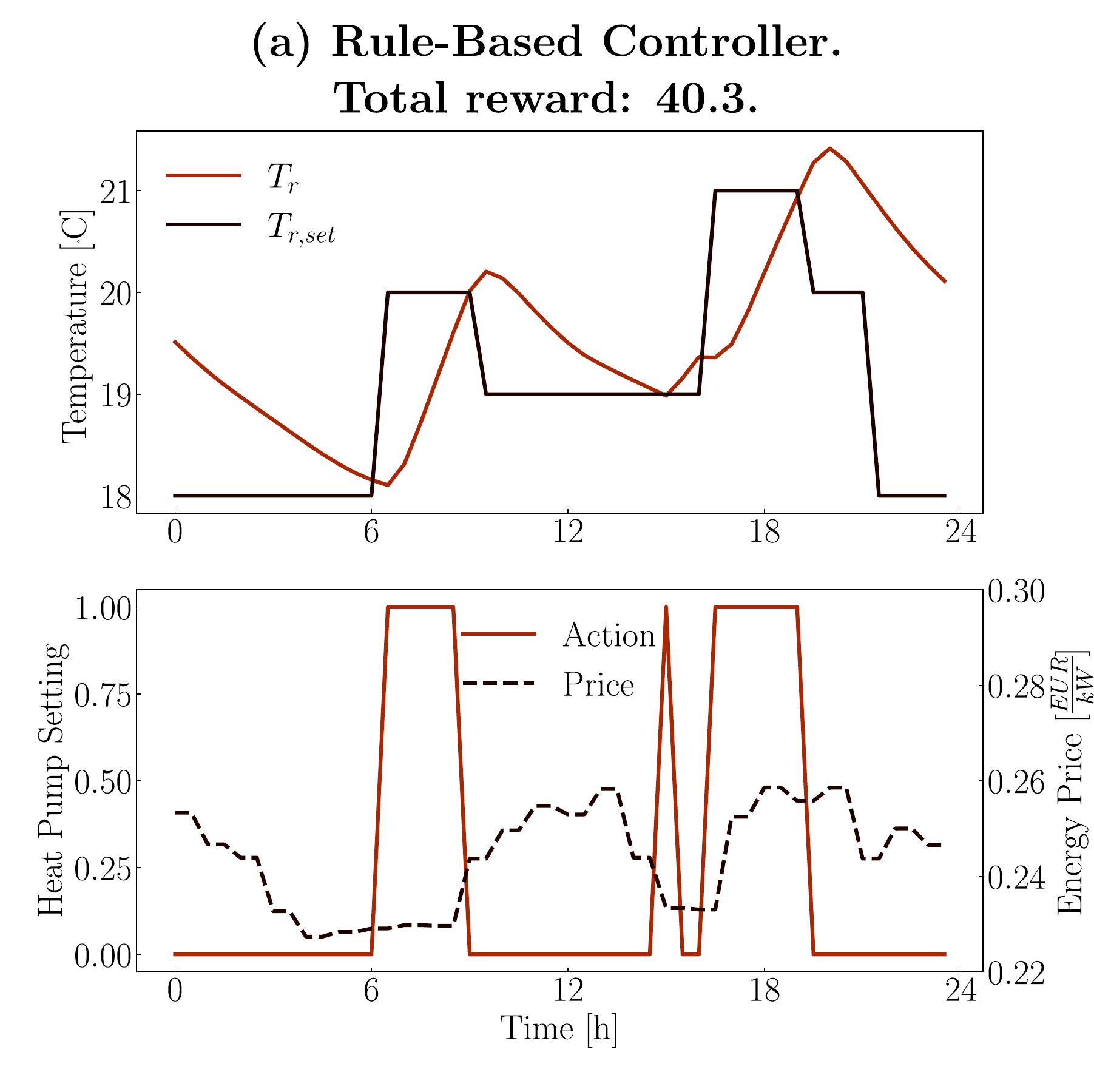}
    \includegraphics[width=0.45\textwidth]{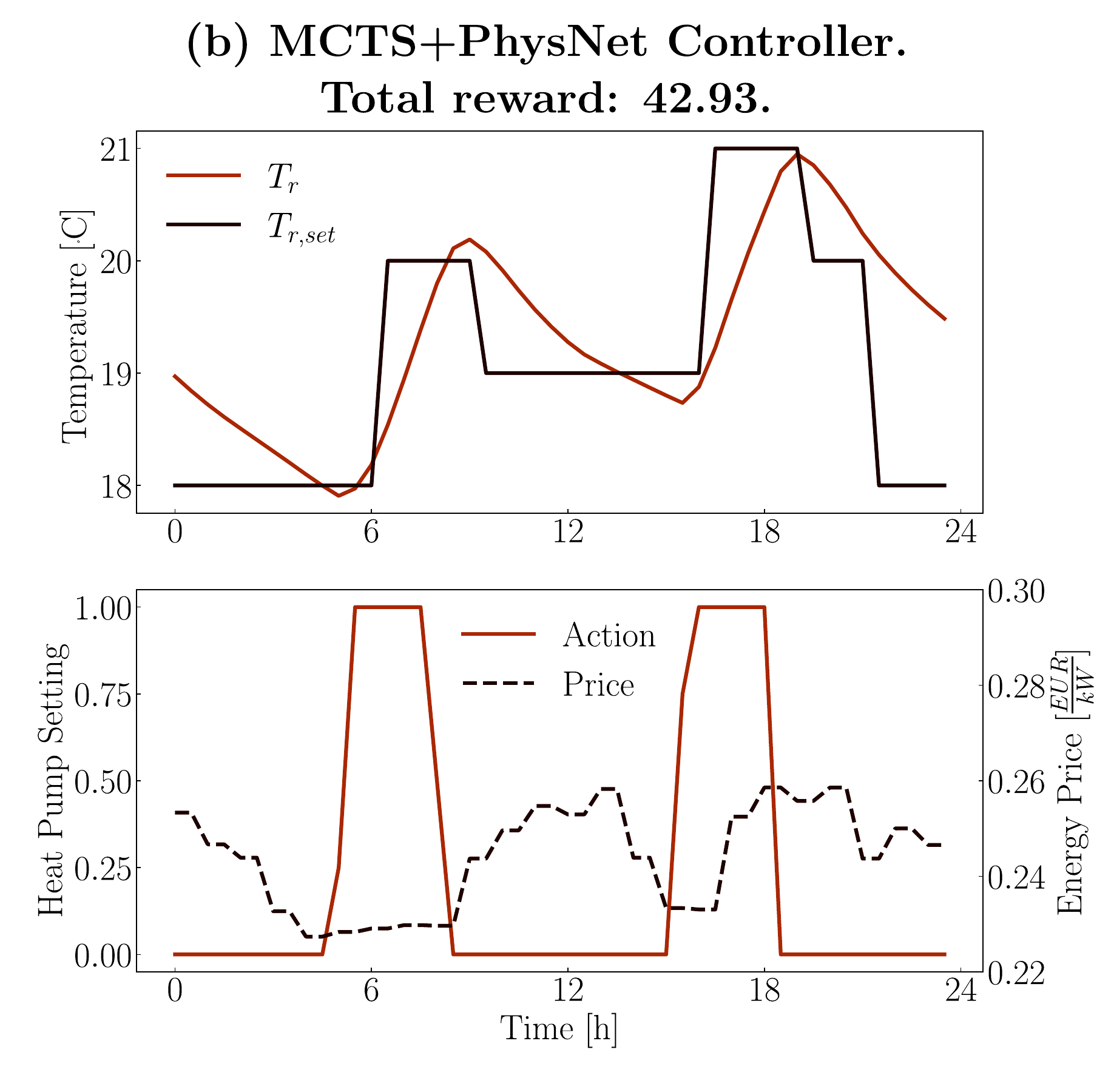}
    \caption{Action sequence selected by (a)~the rule-based bang-bang policy and (b)~the PhysNet-based Vanilla MCTS respectively, for the same randomly selected day.
    The MCTS policy manages to stay closer to the desired temperature, while heating the room at times with lower prices, thus attaining a higher reward compared to the rule-based controller.}
    \label{fig:policies}
\end{figure}

Finally, to achieve more intuitive insight into how our best control policy of MCTS+PhysNet behaves, \cref{fig:policies} compares its decisions over a randomly selected test day with the rule-based baseline.
We note that the rule-based controller tends to over-heat the room (\eg at times $\sim$10:00 and $\sim$20:00), obviously without any consideration of the energy prices.
Conversely, MCTS achieves to better align the room temperature with its desired setpoint. It also manages to exploit low-price points by pre-heating the room (\eg around $\sim$6:00 and $\sim$16:00).
Consequently, MCTS achieves higher rewards, as discussed previously.

\subsection{Efficient tree search with AlphaZero MCTS}
\label{sec:results_alphazero}
A drawback of the Vanilla MCTS algorithm is its 
high computational cost to take a control action, stemming from simulating many rollouts.
As has been shown 
for problems of game playing~\cite{silver2018general, schrittwieser2020mastering}, the addition of prior knowledge into the tree search can increase its performance in terms of managing to achieve higher rewards using fewer simulations. 
As an example of such an improved MCTS algorithm, we implemented the AlphaZero MCTS algorithm (based on~\cite{silver2017mastering}, see \cref{sec:method-mcts}).
To quantify the computational cost required by MCTS to obtain each action, we use the number of simulations as a metric (\cf a higher number of simulations implies a higher computational cost).
The NN used for the prior probability $P(\tilde{x}^k, \tilde{u})$ in \cref{eq:alphazero_uct} is trained offline at the end of each day by using simulated samples obtained with data from previous days.
These samples are obtained by using the Vanilla MCTS algorithm and by searching and evaluating on the same simulated environment. 
Since that training is performed offline, it can be done at times where a smart controller is typically not required (\eg at night).
Given the offline nature, in collecting the training sample trajectories, we 
may use a high number of simulations for the Vanilla MCTS to guide them, and thus obtain high-quality experience samples to train the prior-policy NN.

\begin{figure}
    \centering
    \includegraphics[width=0.45\textwidth]{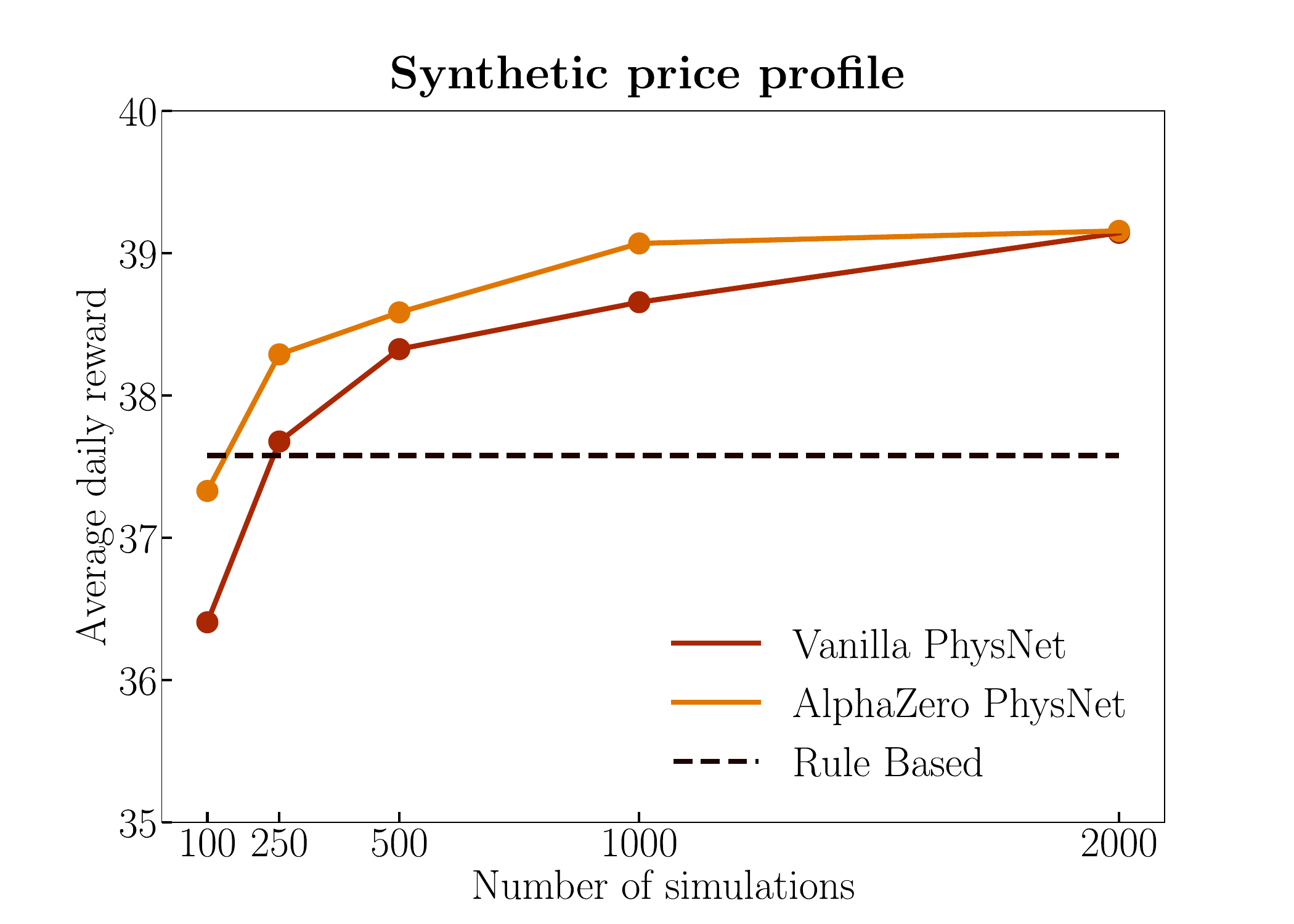}
    \includegraphics[width=0.45\textwidth]{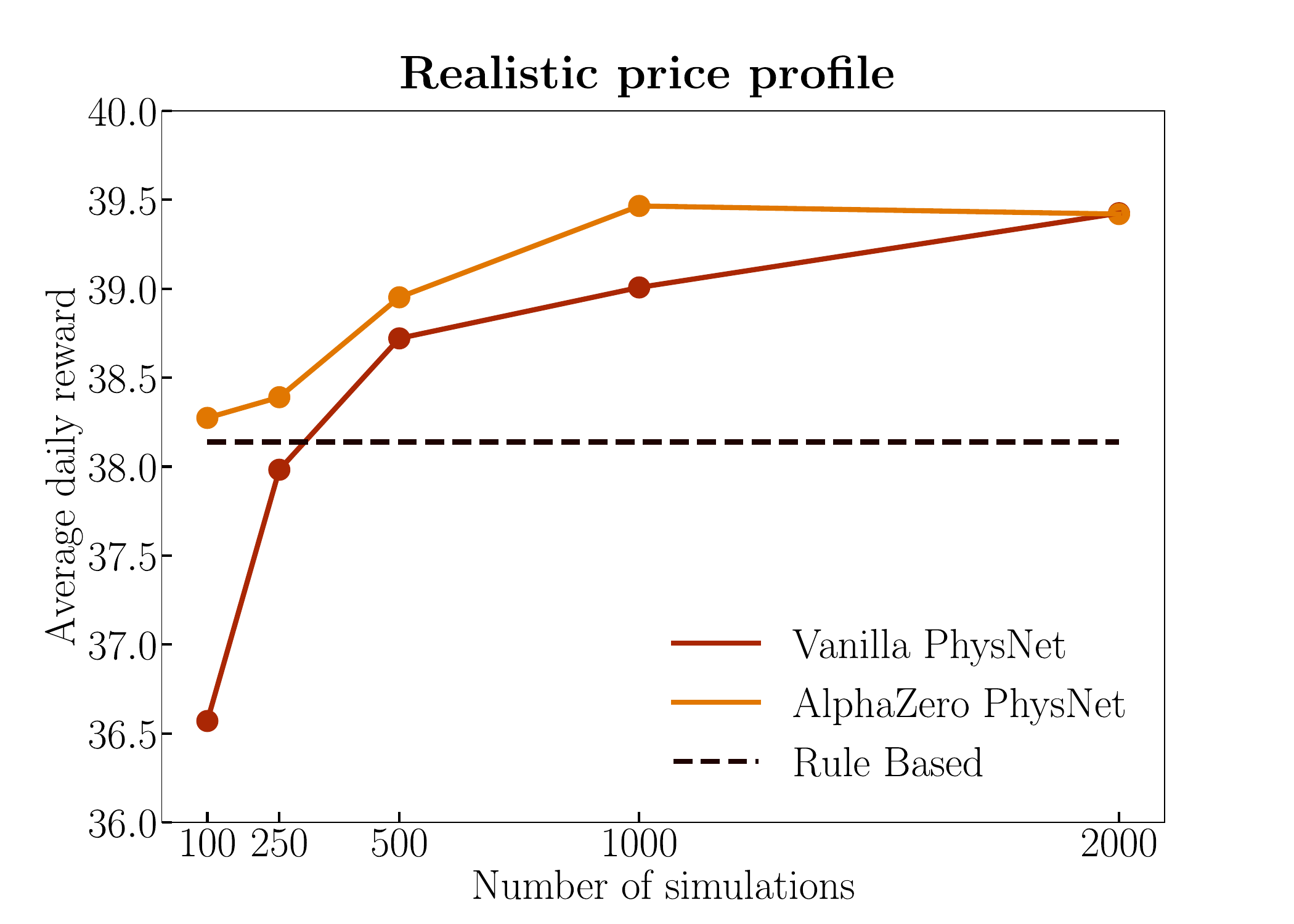}
    \caption{Average daily rewards obtained by the Vanilla MCTS and AlphaZero MCTS when applied to 11 test days with a PhysNet simulator. The addition of a prior-policy NN (\ie the AlphaZero algorithm) achieves a substantial increment in rewards when a low number of simulations is performed.}
    \label{fig:az_vs_van_results}
\end{figure}

We compare the results of the Vanilla MCTS algorithm with AlphaZero MCTS, 
using the same approach as \cref{sec:results_mcts}.
\Cref{fig:az_vs_van_results} shows the 
average daily reward obtained 
for a varying number of simulations. 
We adopt the PhysNet model for these simulations in MCTS, since we previously assessed its performance to be superior to its black-box counterpart.
When a low number of simulations is performed, AlphaZero achieves 
a notably higher reward (2\% higher on average for $\le$500 simulations), both for the synthetic and the realistic price scenarios compared to the Vanilla version.
When increasing the number of MCTS simulations, this benefit continues to hold but diminishes until AlphaZero and Vanilla MCTS ultimately converge (since this implies the large number of unrolled tree scenarios for Vanilla MCTS also suffice to make a (near-)optimal decision).
From a practical perspective, we conclude that for real-world applications the AlphaZero MCTS controller is to be preferred since it efficiently finds valid control actions with a low number of simulations.

\section{Final Discussion}

\subsection{Conclusions}
\label{sec:conclusions}

In this paper, we presented an MCTS algorithm with a PiNN applied to a residential heating Demand Response problem.
We first deployed and expanded PhysNet, a previously proposed PiNN, allowing it to produce multi-time-step predictions of room temperature and energy consumption of a heat pump.
After establishing PhysNets predictive performance (32\% lower MAE prediction error in a low training data regime, compared to a black-box model),
we use it as a simulator for the MCTS control algorithm.
Compared to a rule-based controller baseline, we noted a substantial improvement of a Vanilla MCTS controller using PhysNet in terms of the considered reward function (up to 9\% reduction in average cost and 19\% increase in thermal comfort).
Subsequently, we showed how adopting the more advanced AlphaZero MCTS helps to achieve such benefits at lower computational cost (\ie fewer tree rollout simulations), by incorporating a policy NN to guide action explorations more efficiently.

Thanks to its flexibility to incorporate state-dependent constraints on the allowed actions, as well as its planning nature, we believe MCTS is a valid technique for Demand Response control of a residential heater.
Moreover, compared to conventional RL approaches, MCTS offers more interpretable control actions (\eg one could inspect the various rollouts in the tree constructed to compare effects of other possible action trajectories). 
Such higher degree of interpretability (and the failure case analysis it enables) could spur higher levels of trust and adoption in real-world commercial settings.
Conceptually, MCTS creates
a bridge between model-based techniques (such as MPC) and model-free data-driven ones (such as RL), 
exploiting benefits from both (interpretability from MPC, efficient and data-driven learning from RL).

We hope that our exploration of MCTS (with our contributions of incorporating PiNN and analysis of AlphaZero-like policy network) spurs further research specifically on its application in residential energy control algorithms.
Next we list current limitations (\cref{sec:concl-limitations}) and directions for such future work (\cref{sec:concl-future}) in this area.

\subsection{Strengths and Limitations}
\label{sec:concl-limitations}
Although our methods provide several benefits, different limitations need to be taken into consideration when applied in real-world settings.
Regarding the application of PhysNet, our proposed PiNN model, we showed the benefits of infusing physical knowledge, both for forecasting and control performance.
PhysNet relied on estimating an interpretable latent state variable, \ie the building mass temperature.
However, this was achieved by a (simplified) RC model of the building, which requires proper initialization to work properly.
The latter implies that in practice, it would call for numerical optimization or empirical estimation, which incurs a potential issue in deploying it at scale across a multitude of households.
Furthermore, we note that the estimated hidden state may not accurately represent the actual building mass temperature (\cf the training objective only included room temperature prediction accuracy) --- thus it cannot be reliably used as an estimation thereof.

Regarding MCTS, a core characteristic is its reliance on (reasonably accurate) simulation of action sequences and their effects on the system state.
On the plus side, such explicit rollouts allow to naturally account for (possibly state-dependent) constraints, since impermissible actions can be easily pruned from the tree.
Further, the constructed tree facilitates interpretability (see before, \cref{sec:conclusions}).
Yet, a main drawback of MCTS is its computational cost at inference time, 
since the action decision is based on building the tree dynamically from the current system state onwards.
Given that building the tree is based on Monte Carlo simulations, to make (near-)optimal decisions a potentially large number of simulations is required.
The latter can be prohibitive in residential applications such as the considered heat pump case, either or both because of limited computational requirements or real-time nature (\eg if the control timestep would be far lower than the currently considered 30\,min intervals).
Although more efficient tree expansion strategies (\cf the AlphaZero approach we applied) partially address this, real-world deployment may require further improvements.


\subsection{Future Work}
\label{sec:concl-future}
Following our study offering (some of) the first steps into MCTS solutions for residential building energy management systems (BEMS), 
further research is needed to push the technology towards large-scale real-world adoption.
Although our results already confirmed the potential of MCTS in BEMS, a comprehensive benchmarking against competing state-of-the-art control algorithms (including both MPC and RL approaches) in terms of control performance, computational cost, and scalability should be carried out to robustly establish MCTS's competitiveness.
Further refining of the MCTS algorithm based on recent evolutions in the field (\eg MuZero~\cite{schrittwieser2020mastering}) is another direction of future work.
These innovations could further improve the technique's performance by, for example, merging the modeling and planning tasks in a single solution within the tree search.
Besides methodological improvements, the exploration of MCTS in terms of application scenarios is a valuable research direction.
For example, in scenarios with non-stationary constraints --- \eg they strongly vary dynamically over time --- and/or more complex ones than what we considered in this paper, MCTS-like solutions could be particularly beneficial.
Finally, actual field trials with real-world deployment of MCTS controllers should be carried out to establish their practical applicability.

\section*{Declaration of Competing Interest}
The authors declare that they have no known competing financial interests or personal relationships that could have appeared to influence the work reported in this paper.

\section*{Acknowledgments and funding}
This work was supported in part by the European Union’s Horizon 2020 research and innovation programme under the projects BRIGHT (grant
agreement no. 957816) and BIGG (grant agreement no. 957047). 

We also thank Marie-Sophie Verwee for her technical support in the deployment of our work.

\section*{Declaration of generative AI and AI-assisted technologies in the writing process}
During the preparation of this work the author(s) used ChatGPT 3.5 in order to enhance the language quality of the manuscript. After using this tool/service, the author(s) reviewed and edited the content as needed and take(s) full responsibility for the content of the publication.

\printcredits

\bibliography{bibl}

\appendix

\section{Hyperparameters}
\label{appendix:hyperparameters}
The PhysNet network is composed of two separate Feed Forward Neural Networks (FFNNs), the encoder and the predictor. The encoder has a single hidden layer composed of 32 neurons activated with a ReLU function. Its output layer is activated with a tanh function. The predictor has a single hidden layer composed of 64 neurons activated with a ReLU function. The output of the predictor layer has two different activation functions: tanh for the temperature, and ReLU for the energy consumption. The past observations depth value $d$ is fixed to 24 time-steps (12 hours). All the neural network parameters are trained with an Adam optimizer.
The prior policy neural network~$p(\tilde{x})$ of the AlphaZero algorithm is a FFNN with 2 hidden layers of 64 and 32 neurons activated with a ReLU function. 
Its final output gets activated through a softmax function and the whole network gets trained with a Cross-Entropy loss and an Adam optimizer.
All the neural networks used in this paper were implemented using PyTorch 2.0~\cite{pytorch}.

\section{Rule Based Controllers}
\label{appendix:rulebased_controller}
The rule-based bang-bang controller referred to in this work heats the room whenever its temperature drops below the desired one. Mathematically, its policy is defined as:
\begin{equation}
    u_t = 
    \begin{cases}
      1 & \text{if} \;\; T_{\text{r}, t} < T_{\text{r\_set},t} \\
      0 & \text{otherwise}
    \end{cases}
\end{equation}
The discrete rule-based controller is defined as:
\begin{equation}
    u_t = 
    \begin{cases}
      0 & \text{if} \;\; T_{\text{r}, t} > T_{\text{r\_set},t} \\
      0.25 & \text{else if} \;\; T_{\text{r}, t} > T_{\text{r\_set},t} - 0.05 \\
      0.5 & \text{else if} \;\; T_{\text{r}, t} > T_{\text{r\_set},t} - 0.15 \\
      0.75 & \text{else if} \;\; T_{\text{r}, t} > T_{\text{r\_set},t} - 0.25 \\
      1 & \text{otherwise}
    \end{cases}
\end{equation}
Finally, the continuous controller used to generate data in \cref{sec:results_forecasting} is defined as:

\begin{equation}
    u_t = \min\left(2(T_{\text{r\_set}, t} - T_{\text{r}, t})^+, \; 1\right)
\end{equation}

\section{Cumulative noise}
\label{appendix:cumulative_noise}
To evaluate our experiments more realistically, we added random noise to the future outside temperature. The added noise will simulate the prediction error of a realistic forecaster tool.

For each time-step, we sample from a Bernoulli distribution to determine the sign of the added noise (positive or negative). We then sampled and added random noise from a normal distribution (with the previously decided sign) to the outside temperature. The noise added gets cumulated while considering values more far into the future. Each addition get multiplied with a $\sigma$ (or $-\sigma$) hyperparameter that determines the magnitude of the added noise

\section{Normalization of the rewards}
\label{appendix:normalization_rewards}
To use the rewards obtained from \cref{eq:reward_funct} in our MCTS algorithm, we scaled the values into the closed unit interval. To do so, we used the standard min-max equation:
\begin{equation}
    \rho_{\text{norm}} \doteq \frac{\rho - \rho_{\text{min}}}{\rho_{\text{max}} - \rho_{\text{min}}} \;\; ,
\end{equation}
where $\rho_{\text{norm}} \in [0, 1]$ is the scaled value of $\rho \in [\rho_{\text{min}}, \rho_{\text{max}}]$.
To use this formula, we need to define $\rho_{\text{min}}, \rho_{\text{max}}$ as domain bound for $\rho$. Given the reward formula (\cref{eq:reward_funct2}), the upper bound is easily defined as $\rho_{\text{max}} \doteq 0$, corresponding to the scenario where $\uphys = 0$ and $\Tr = T_{\text{r\_set}}$. Defining the lower bound $\rho_{\text{min}}$ requires instead an empirical choice, as in principle the rewards are not limited to a finite interval. To define $\rho_{\text{min}}$, we picked the two lower bounds of each part of the reward function, namely the cost optimization and the thermal comfort optimization. For the cost minimum value, we considered the highest cost multiplied by the highest heat pump energy consumption measured in our train set. For the thermal comfort minimum, we picked the worst case as a difference between $\Tr$ and $T_{\text{r\_set}}$ of $2\text{\centigrades }$ multiplied by $c_1$.

\end{document}